\documentclass[12pt]{article}
\usepackage{graphicx}
\usepackage[margin=0.7in,a4paper]{geometry}
\usepackage{setspace}
\usepackage[T1]{fontenc}
\usepackage[utf8]{inputenc}
\usepackage{amsmath,amssymb,amsfonts}
\usepackage{multirow,booktabs,tabularx,longtable}
\usepackage[table,xcdraw]{xcolor}
\usepackage{booktabs}
\usepackage{tabularx}
\usepackage{longtable}
\usepackage{caption}
\usepackage{algorithm}
\usepackage{algpseudocode}
\usepackage{graphicx}
\usepackage{titlesec}
\usepackage{float}
\usepackage{placeins}   
\usepackage{subcaption}
\usepackage{authblk}
\usepackage[authoryear]{natbib}
\usepackage{hyperref}
\usepackage{xcolor}
\hypersetup{
    colorlinks=true,
    citecolor=blue,
    linkcolor=blue,
    urlcolor=blue
}
\usepackage{bm}

\title{A Copula-Based Regression Framework for Enhanced Prediction under Heteroscedasticity}

\author{
D. K. Hemachandra$^{1^*}$,
S.G.J. Senarathne$^{1,2}$,
M. B. Dehideniya$^{2}$\\
{\small $^{1}$Postgraduate Institute of Science, University of Peradeniya, Peradeniya, Sri Lanka}\\
{\small $^{2}$Department of Computer Science and Statistics, University of Peradeniya, Peradeniya, Sri Lanka}
}

\date{}

\begin{document}

\maketitle

\begin{abstract}
Classical regression approaches, including ordinary least squares, rely on strong assumptions such as constant variance and normality of residuals, which are often violated in real-world data. Although log-transformation is commonly used to stabilise variance, it may introduce re-transformation bias and fail to address heteroscedasticity and asymmetric dependence structures adequately.
To overcome these limitations, this study proposes a copula-based regression framework for modelling data in the presence of heteroscedastic error structures. The proposed copula-based regression framework separates marginal distributions of the response and explanatory variables from their dependence structure, allowing flexible modelling of different tail-dependent relationships. The proposed approach explicitly accounts for heteroscedasticity without requiring restrictive distributional assumptions.
A comprehensive simulation study and two real-world applications were considered under heteroscedastic scenarios to compare the performance of proposed method over existing methods including classical regression, log-linear, quantile regression models and generalized additive model for location, scale, and shape(GAMLSS). The simulation results demonstrated that the proposed copula-based model consistently outperformed conventional approaches, achieving an average mean absolute percentage error of 0.21, compared with 0.27 and 0.36 for the linear and log-linear models, respectively. Across all simulation settings, the copula-based model reduced prediction errors by approximately 6\%–33\% relative to the linear model and 24\%–57\% relative to the log-linear model.  In the first application, which exhibited clear heteroscedasticity, the copula-based model achieved the lowest MAPE, although the overall differences between the copula-based and GAMLSS models were not substantial.  In the second application, all models achieved similarly low predictive performance, as there was a moderate linear relationship between the response and predictor variables. Together, the two real-world applications confirmed the effectiveness of the proposed model under both favourable and less favourable conditions. Overall, the findings indicate that no single model consistently dominates across all settings, while copula-based regression provides a flexible and competitive alternative for heteroscedastic data.


\end{abstract}
\clearpage

\section{Introduction}

Regression analysis is one of the most widely used statistical approach for modelling the relationship between a response variable and one or more explanatory variables, with applications in many fields including economics, finance, insurance, health sciences, and engineering. Such models often use as Ordinary Least Squares (OLS) to estimate the model parameters and are popular due to their simplicity and interpretability. However these models, rely on strong assumptions, including linearity, normally distributed errors, and constant error variance (homoscedasticity). In many real-world applications, these assumptions are often violated.

A particularly important violation of the constant variance assumption is \emph{heteroscedasticity}, where the variance of the regression errors varies with the level of the covariates or the conditional mean. Heteroscedasticity is common in real world data including financial returns, insurance claims, biomedical data, and socio-economic studies, where variability often increases with scale or risk exposure \citep{Wooldridge2019, duan1998retransformation}. When heteroscedasticity is present, OLS estimators remain unbiased but are no longer efficient, and standard errors become inconsistent, leading to invalid inference and unreliable prediction intervals.

Several traditional remedies have been proposed to address non-constant error variance in regression. Variance-stabilising transformations may reduce heteroscedasticity but complicate interpretation and often fail to capture complex dependence structures. However, log-transformations are not always effective in addressing heteroscedasticity and may introduce additional complications, particularly when predictions are re-transformed back to the original scale. As shown by \citet{duan1998retransformation, manning2001logmodels}, re-transformation can lead to biased estimates when error variance is non-constant. Moreover, log-linear models may fail to adequately capture asymmetric or tail-dependent relationships commonly observed in real-world data \citep{manning2001logmodels, duan1998retransformation, Kolev2009}. Weighted Least Squares improves efficiency by incorporating a variance function, but its performance depends critically on correct specification of the variance structure, which is rarely known in practice \citep{Wooldridge2019}. Generalised Linear Models (GLMs) allow for non-normal response distributions and mean--variance relationships; however, they still rely on restrictive distributional assumptions and predefined variance structures, which may limit their ability to adequately capture skewed, heavy-tailed, or asymmetrically dependent data structures \citep{Wooldridge2019, Stasinopoulos2024, Rigby2020}. As noted by \citet{Parsa2011, Kolev2009}, these conventional approaches may break down when residuals are both non-normal and heteroscedastic.

Quantile regression (QR) provides a flexible approach to modelling heteroscedasticity by characterising the conditional distribution of the response through multiple conditional quantiles rather than only the conditional mean. Building on this framework, \citet{Wang2023} proposed a penalised multiple quantile regression approach for identifying heteroscedasticity and selecting influential variables in high-dimensional regression models. Their simulation studies demonstrated improvements in estimation accuracy and variable selection under heteroscedastic conditions compared with existing penalised quantile regression methods.
Alternately Generalized Additive Models for Location, Scale and Shape (GAMLSS) framework extends conventional regression by modelling multiple parameters of the conditional distribution. Specifically, GAMLSS allows not only the location parameter but also the scale, skewness, and kurtosis parameters of the response distribution to depend on explanatory variables, thereby providing a flexible mechanism for modelling non-constant variance and other distributional characteristics \citep{Rigby2005,Stasinopoulos2024,Correa2023}. This flexibility enables GAMLSS to estimate multiple aspects of the conditional distribution and accommodate complex heteroscedastic patterns. 
Although quantile regression and GAMLSS provide effective strategies for modelling heteroscedasticity through conditional quantiles and distributional parameters, respectively, their primary emphasis has been on distributional modelling, estimation, or variable selection rather than explicitly modelling dependence structures or systematically evaluating predictive performance across alternative regression frameworks.

Copula-based regression models provide a flexible alternative by allowing the dependence structure between variables to be modelled separately from their marginal distributions. This separation enables the modelling of non-linear, asymmetric, and tail dependence without imposing restrictive assumptions on the marginals. Early work by \citet{Parsa2011} demonstrated the potential of copula regression as an alternative to OLS and GLM, while \citet{Kolev2009} provided a comprehensive survey of copula-based regression frameworks and dependence structures. Subsequent extensions include skew-normal copula regression to capture asymmetric dependence \citep{Wei2018}, high-dimensional copula regression methods \citep{Cai2021}, and non-parametric Bayesian copula estimation using empirical Bernstein and checkerboard copulas \citep{ Lu2023}. Empirical evidence suggests that copula regression can provide a better model fit than conventional regression approaches when analysing correlated response variables, as it explicitly models the dependence structure between outcomes \citep{Lee2020}.

Despite the considerable progress in copula-based regression, existing approaches have primarily focused on modelling complex dependence structures rather than explicitly addressing heteroscedastic error structures in regression settings \citep{Kolev2009, Parsa2011, Wei2018, Cai2021}. Consequently, the potential of copula-based methods to accommodate non-constant variance while preserving flexible dependence modelling remains insufficiently explored. This limitation is particularly relevant in applications such as finance, insurance, and economics, where heteroscedasticity is frequently observed \citep{Wooldridge2019}. Furthermore, systematic comparisons between copula-based and conventional regression approaches under heteroscedastic conditions are scarce, particularly with respect to parameter estimation, uncertainty quantification, and predictive performance. These gaps highlight the need for a flexible regression framework that can simultaneously account for heteroscedasticity and complex dependence structures.

This study addresses these gaps by developing copula-based regression models that explicitly account for non-constant error variance. Simulation studies and real-data applications are used to evaluate the performance of the proposed models relative to conventional linear regression approaches, with particular emphasis on estimation accuracy, interval estimation, and prediction under heteroscedasticity.
The main contributions of this work are threefold: (i) introducing a methodological framework for incorporating heteroscedasticity within copula regression, (ii) providing a systematic comparison between copula-based and classical regression models, and (iii) offering practical guidance on when copula regression yields more reliable inference and improved prediction.

\begin{figure}[htbp]
\centering
\includegraphics[width=0.8\linewidth]{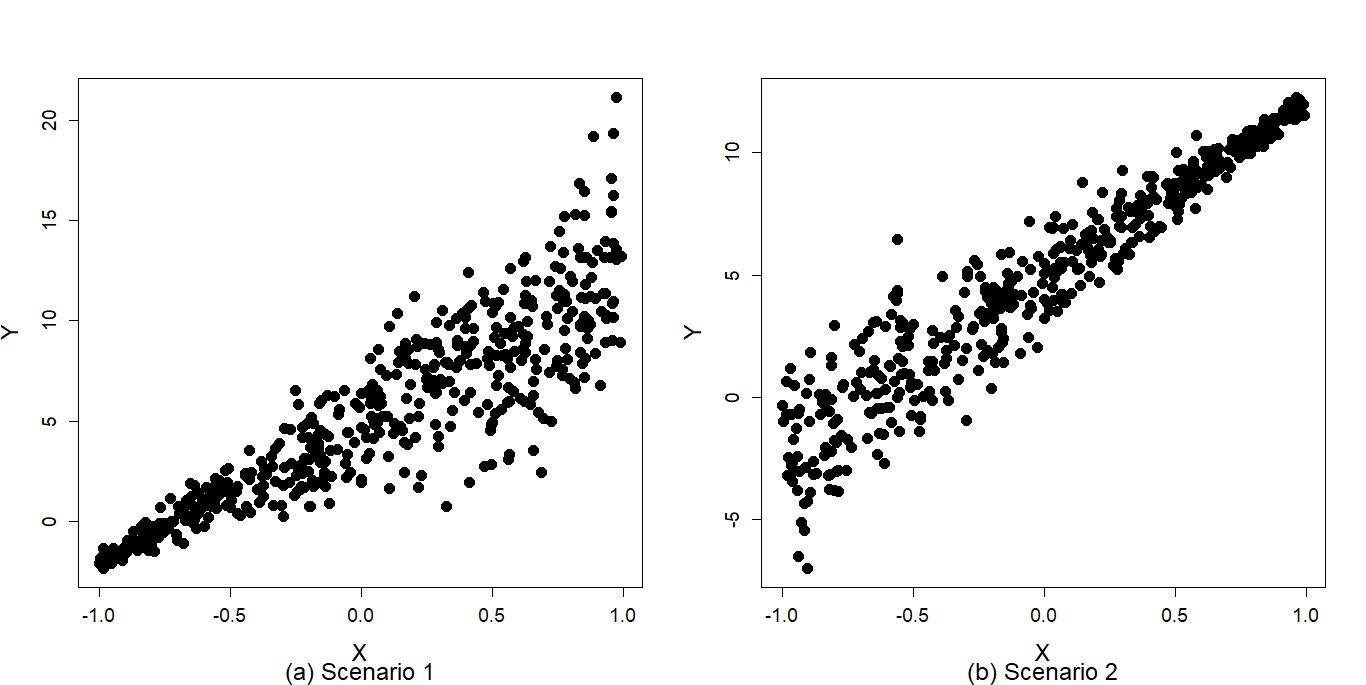}
\caption{Two scenarios of non-constant error variance in linear regression}
\label{fig:heteroskedasticity_scenarios}
\end{figure}

Figure~\ref{fig:heteroskedasticity_scenarios} illustrates two representative patterns of non-constant error variance commonly observed in regression settings. In the first Scenario, the variability of the response increases in the upper tail of the distribution, leading to a fan-shaped pattern where high values of the response exhibit substantially larger dispersion. This behaviour is frequently encountered in financial and insurance data, where risk and uncertainty grow with scale. The second Scenario corresponds to increasing variability in the lower tail, where dispersion is more pronounced for smaller values of the response. This pattern arises in applications involving loss data or constrained outcomes, and it violates the constant variance assumption underlying classical regression models.

\FloatBarrier

These two Scenarios highlight that heteroscedasticity is not limited to monotonic variance trends but may depend on the entire conditional distribution of the response. Traditional alternative approaches such as Weighted Least Squares and GLMs rely on correct specification of the variance or mean--variance relationship and typically assume symmetric dependence. Consequently, they are unable to accommodate tail-specific variability, asymmetric dependence, or non-linear association structures. The copula-based framework proposed in this study overcomes these limitations by explicitly modelling the dependence structure between the response and explanatory variables, allowing heteroscedasticity and tail behaviour to be captured flexibly without restrictive distributional assumptions.

The remainder of this paper is organised as follows. Section~\ref{Methodology} presents the proposed methodology for constructing a copula-based regression model using a bivariate copula. Sections~\ref{Simulation Study} and ~\ref{Application} report the results of simulation studies and real-data applications. Finally, Section~\ref{Conclusion} concludes the paper.

\section{Methodology}
\label{Methodology}

\subsection{Classical Regression Framework}
\label{sec:classical_regression}

Let $Y$ denote a response variable and $\mathbf{X} = (X_1, X_2, \ldots, X_p)$ a set of explanatory variables, $\beta_0, \beta_1, \ldots, \beta_p$ be unknown regression coefficients, and $\varepsilon$ be the random error term. In the classical linear regression framework, the relationship between $Y$ and $\mathbf{X}$ is given by
\begin{equation}
Y = \beta_0 + \beta_1 X_1 + \cdots + \beta_p X_p + \varepsilon,
\end{equation}
where $\varepsilon \sim N(0, \sigma^2)$ with mean zero and constant variance \(\sigma^2\). 

A key assumption of classical regression is the homoscedasticity of error terms. When the error variance depends on the covariates, the model exhibits heteroscedasticity. In such settings, OLS estimators are inefficient and standard errors are inconsistent, resulting in unreliable inference and prediction.

\subsection{Copula-Based Regression}
\label{sec:copula_regression}

Copula-based regression provides a flexible framework for modelling dependence between the response and explanatory variables without restrictive assumptions on their joint distributions. Let $F_Y(y)$ denote the marginal distribution of $Y$ and $F_{X_j}(x_j)$ the marginal distribution of $X_j$, for $j = 1, \ldots, p$. By Sklar's theorem \citep{Sungur2005,Kolev2009}, there exists a copula $C$ with a copula parameter $\bm{\theta}$ such that
\begin{equation}
H(y, x_1, \ldots, x_p) =
C\big(F_Y(y), F_{X_1}(x_1), \ldots, F_{X_p}(x_p); \bm{\theta}\big),
\label{Eq:skalr1}
\end{equation}
where $H$ denotes the joint distribution of $(Y, \mathbf{X})$. This representation allows the variability of the response variable, including heteroscedasticity, to be modelled separately from the dependence structure. 

Most existing copula-based regression approaches rely on multivariate copula constructions to model the joint distribution of the response and explanatory variables. However, as the number of predictors increases, the number of parameters to be estimated also increases, leading to greater model complexity and computational burden. In contrast, the present work proposes a bivariate copula‑based regression model, which explicitly links the linear predictor $Z = \gamma_0 + \gamma_1 X_1 + \cdots + \gamma_p X_p$ and the response $Y$ through a single bivariate copula, keeping the parameter space relatively simple while capturing key dependence features.

\subsection{Proposed Copula-Based Framework}  

Let consider the same $Y$ and $\mathbf{X}$ defined in Section~\ref{sec:classical_regression} and linear predictor  $Z$ defined in Section~\ref{sec:copula_regression}. Let $F_Y(y)$ and $F_Z(z)$ denote the cumulative distribution functions of $Y$ and $Z$, respectively. According to Equation \eqref{Eq:skalr1}, their joint distribution can be expressed as
\begin{equation}
F_{Y,Z}(y,z)=C\big(F_Y(y),F_Z(z); \bm{\theta}\big).
\end{equation} 
Defining $u = F_Y(y)$ and $v = F_Z(z)$ transforms the marginal distributions into standard uniform variables on the interval $(0,1)$. Using the alternative representation of Sklar's theorem, the bivariate copula function can be expressed in terms of the inverse marginal distribution functions as follows. 
\begin{equation}
C(u,v;\boldsymbol{\theta}) = F_{Y,Z}\left(F_Y^{-1}(u),\, F_Z^{-1}(v)\right), 
\qquad 0 < u, v < 1
\end{equation}
As such, the proposed copula-based framework models the dependence structure between the response variable and the linear predictor using a bivariate copula.

Selecting an appropriate bivariate copula to model dependence can be challenging, as 
there are multiple copulas which can be used to model similar tail behaviours. In this context, Archimedean copulas are particularly attractive due to their simple mathematical form, computational efficiency, and flexibility in capturing a wide range of dependence structures using a single copula parameter. They also provide a direct and interpretable representation of dependence through correlation parameters such as Kendall's tau and Spearman rank correlation, while minimising the complexity associated with multivariate copula constructions. For these reasons, Archimedean copulas offer an effective framework for modelling dependence in regression settings.

The parameters of the proposed copula-based regression model are estimated using the Maximum Likelihood Estimation (MLE) approach. Based on the joint distribution of the response variable $Y$ and the linear predictor $Z$, the joint likelihood function is constructed using the copula representation. This can be decomposed into marginal and dependence components, allowing the likelihood to be expressed in terms of the copula density and the marginal distributions. In particular, the conditional likelihood of $Y$ given $Z$ is derived from the joint copula structure, enabling estimation of the model parameters by maximizing the likelihood with respect to $Y$ for given values of the covariates $X$. The conditional distribution required for this formulation is obtained as follows:

\begin{equation} 
C_{Y \mid Z}(y, z \mid \bm{\theta)} = \frac{\partial C(u, v \mid \bm{\theta)})}{\partial u}
\label{eq:8}
\end{equation} 

The overall parameter estimation framework for the proposed model is summarized in the following algorithm.

\begin{algorithm}[htbp]
\caption{Copula-Based Regression Framework with Linear Model Initialization}
\label{alg:copula_updated}
\begin{algorithmic}

\State \textbf{Input:} Sample Dataset 
$D_{\text{sample}} = \{(x_{1i}, x_{2i}, \ldots, x_{pi}, y_i)\}_{i=1}^n$
\Statex
\State \textbf{Step 1:} Define a new random variable
\begin{equation}
{Z} = {\gamma}_0 + {\gamma}_1 X_1 + {\gamma}_2 X_2 + \ldots + {\gamma}_p X_p
\label{eq:6}
\end{equation}     
\Statex
\State \textbf{Step 2:} Approximate the marginal distribution of $ Z \sim f_z(\bm{\alpha})$ using empirical distribution technique and define the distribution of $Y \sim \mathcal{N}({Z}, \sigma^2)$

\Statex
\State \textbf{Step 3:} Apply the probability integral transform to both marginal distributions to map the variables onto the unit interval. Accordingly, define
\begin{equation}
U = F_Y(y), \quad V = F_Z(z),
\label{eq:7}
\end{equation}
Select a suitable Archimedean bivariate copula to approximate the joint distribution of $U$ and $V$, and define the conditional density
\begin{equation}
F_{U \mid V}(u \mid v)
=
C_{U \mid V}(u,v \mid \boldsymbol{\theta})
=
\frac{\partial C(u,v;\boldsymbol{\theta})}{\partial u}.
\label{eq:8b}
\end{equation}    
\Statex
\State \textbf{Step 4:} Estimate the parameters of the model including $(\gamma_0 , \gamma_1 , \gamma_2 , \ldots , \gamma_p, \boldsymbol{\alpha},\sigma^2,\bm{\theta})$ using maximum likelihood technique
\Statex

\end{algorithmic}
\end{algorithm}
\FloatBarrier

The proposed copula-based regression framework integrates classical linear regression with copula theory to flexibly model the relationship between \(\mathbf{X}\) and \(Y\) under heteroscedasticity. The procedure consists of the following steps.

First, an observed dataset consisting of a response variable $(Y)$ and a collection of explanatory variables $(X_1, X_2, \ldots, X_p)$ is considered. It is assumed that all retained explanatory variables are statistically significant. Based on these significant predictors, a final linear regression model is fitted, and a linear predictor is constructed as Equation \eqref{eq:6}. The fitted linear predictor $Z$ obtained from the classical regression model is subsequently incorporated into the copula-based regression framework.

In the second step, the regression coefficients $(\gamma_0,\gamma_1,\ldots,\gamma_p)$ are initially estimated using the OLS method. The distribution of the fitted linear predictor $Z$ is then approximated using an empirical distribution approach, and candidate parametric distributions are evaluated using the Akaike Information Criterion (AIC) and Bayesian Information Criterion (BIC). It should be noted that these initial estimates serve only as starting values, as all model parameters are ultimately re-estimated within the copula framework using MLE.

The suitable Archimedean copula family is selected to model the dependence structure between Y and Z (Step 3). The candidate copula families considered in this study include the Clayton and Gumbel copulas, representing different dependence structures and tail dependence behaviours define in Figure~\ref{fig:heteroskedasticity_scenarios}. The optimal copula is selected using model comparison criteria such as AIC and BIC, with the copula yielding the minimum value chosen for subsequent analysis. In this study, the selection is additionally guided by the observed dependence patterns in the simulated data. However, alternative approaches such as other information criteria or goodness-of-fit measures may also be used in practice.

After selecting the copula in Equation \eqref{eq:8b}, all model parameters are estimated using MLE from the joint likelihood based on the selected copula and the corresponding marginal distributions.

\section{Simulation Study}
\label{Simulation Study}

This section presents the results of a simulation study used to evaluate the performance of the proposed copula-based regression model. The results are organised into two parts. First, a simulation study with a single explanatory variable is considered to examine the behaviour of the model under controlled dependence and heteroscedastic error structures. Second, the framework is extended to multiple explanatory variables to assess the robustness of the approach in higher-dimensional settings. 
Accordingly, the simulation study is performed under two distinct Scenarios, denoted as Scenario~1 and Scenario~2, corresponding to the dependence structures illustrated in Figure~\ref{fig:heteroskedasticity_scenarios}. 
For each scenario, three levels of dependence between the linear predictor $Z$ and the response variable $Y$ are considered, corresponding to Kendall's tau ($\tau$) values of 0.60, 0.75, and 0.90 for Cases 1, 2, and 3, respectively. In addition, the predictive performance of the proposed copula-based regression model is compared with those of the classical linear regression and log-linear models across sample sizes of 30, 50, and 100. 
A total of 500 simulation replicates were conducted for each scenario, case, and sample size combination to ensure reliable performance evaluation.

\subsection{Study I - Single Predictor Copula Regression Model}  
\label{sec:case1_single}
This section describes the simulation framework for the case of a single explanatory variable, which serves as the baseline setting for evaluating the proposed copula-based regression model. The explanatory variable $X_1$ is generated from a uniform distribution on the interval $[-1,1]$.

The response variable is constructed through a linear predictor where the true parameter values are fixed at $\gamma_0 = 2$ and $\gamma_1 = 11$. Random noise is introduced through two independent error components: a homoscedastic error $\varepsilon \sim \mathcal{N}(0,\sigma^2)$ and a covariate-dependent error $\varepsilon_x \sim \mathcal{N}(0,\sigma_x^2)$. The variance of the covariate-dependent error is specified as

\begin{equation}
\sigma_x^2 = a_1(X_1 + k_1),
\label{eq:9}
\end{equation}
where the constant $a_1$ controls the magnitude of heteroscedasticity and the parameter $k_1$ determines the direction of the variance pattern, specifically, \(k_1 = 1.1\) corresponds to Scenario~1, whereas \(k_1 = -1.1\) corresponds to Scenario~2 (see Figure~\ref{fig:heteroskedasticity_scenarios}). Let $Z = \gamma_0 + \gamma_1 X_1$ denote the linear predictor. The response variable is then defined as $Y = Z + \varepsilon + \varepsilon_x$, thereby explicitly incorporating non-constant error variance. For each three cases, the parameters values are summarized in Table~\ref{tab:sim_setup_1x}.

\begin{table}[htbp]
\centering
\caption{Simulation setups for one predictor variable}
\label{tab:sim_setup_1x}
\begin{tabular}{lcc}
\toprule
Case & $\sigma$ & $a_1$ \\
\midrule
1 & 4.75 & 4.5 \\
2 & 2.75 & 3.5 \\
3 & 1.00 & 1.5 \\
\bottomrule
\end{tabular}
\end{table}

\FloatBarrier

\begin{figure}[!htbp]
\centering
\includegraphics[width=\linewidth,height=10.5cm]{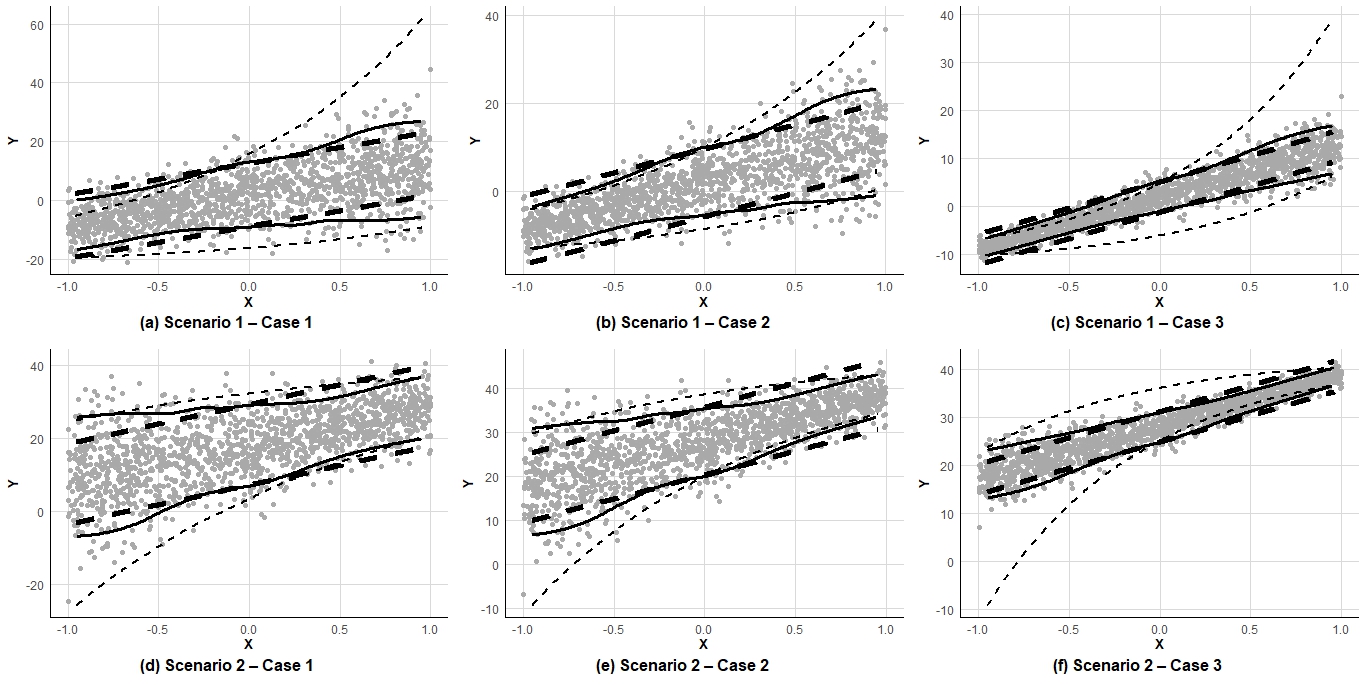}
\centering
\vspace{2mm}  
\includegraphics[width=1\linewidth,height=1.1cm,keepaspectratio]{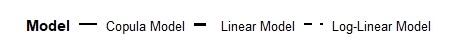}

\caption{95\% prediction intervals for different correlation levels with sample size = 50 }
\label{fig:ci_all_50}
\end{figure}

Figure~\ref{fig:ci_all_50} compares the prediction intervals obtained from the copula-based regression model, log-linear model and the simple linear regression model. Across all simulation configurations, the proposed copula-based model more effectively captures the variability of the response variable, particularly in regions with high residual variability. The prediction intervals obtained from the copula model closely reflect the empirical distribution of the observed data, whereas the classical linear regression model tends to underestimate variability, especially in the tail regions.
The log-linear model also fails to outperform the proposed copula-based approach in terms of predictive performance. In contrast, the copula model more accurately captures the underlying dependence structure and tail behaviour, resulting in prediction intervals that better reflect the observed data patterns. This advantage becomes more pronounced at higher levels of dependence, where the association between the $Z$ and the response variable $Y$ is stronger.

Furthermore, the prediction intervals obtained from the copula-based model provide improved coverage of the observed responses, while the classical linear regression model produces relatively narrow prediction bands that do not adequately adapt to changes in the response variance. Similar patterns are observed for other sample sizes (see Figures~\ref{fig:A1_simple_30} and ~\ref{fig:A2_simple_100} in Appendix ~\ref{app:A})

\FloatBarrier

Table~\ref{tab:mape_results_1} summarizes the Mean Absolute Percentage Error (MAPE) values for the considered modelling approaches across different sample sizes and correlation levels. Across all simulation settings, the copula-based model consistently achieves lower MAPE values than both the classical linear regression and log-linear models, indicating better predictive accuracy. Relative to the linear regression benchmark, the copula model reduces MAPE by approximately 36\% in the first Clayton setting (Case 1, n=30), with larger relative improvements observed in several other settings. The improvement in performance becomes more pronounced as the correlation between $Z$ and $Y$ increases. This indicates that accounting for tail dependence and heteroscedasticity is particularly important when the strength of association is high.

To assess the robustness of the proposed framework, an additional simulation Scenario with a negative regression slope was also considered, and the corresponding results are provided in Appendix ~\ref{app:C}, Table~\ref{tab:mape_results_negative} and Figure~\ref{fig:C4_multi_100}. Changing the regression slope from positive to negative interchanges the Clayton and Gumbel dependence patterns. Nevertheless, the overall conclusions regarding the comparative performance of the proposed copula-based regression model remain unchanged.

\begin{table}[!htbp]
\footnotesize
\centering
\caption{Comparison of Mean Absolute Percentage Error of each models for 500 simulations runs}
\label{tab:mape_results_1}
\renewcommand{\arraystretch}{1.2}
\begin{tabular}{l p{2.6cm} p{1.6cm} p{2.6cm} p{2.6cm} p{2.6cm}}
\toprule
\multirow{2}{*}{Scenario} 
& \multirow{2}{*}{Sample Size} 
& \multirow{2}{*}{Case} 
& \multicolumn{3}{c}{Mean MAPE for Each Model (with standard deviation)} \\ 
\cmidrule(lr){4-6}
 &  &  
& Copula
& Linear
& Log-Linear \\ 
\midrule

\multirow{9}{*}{1}
& \multirow{3}{*}{30}
  & 1 & \textbf{4.26} (0.29) & 6.69 (1.15) & 12.1 (1.50) \\
& & 2 & \textbf{1.53} (0.14) & 3.72 (0.45) & 6.36 (0.60) \\
& & 3 & \textbf{1.48} (0.02) & 2.21 (0.21) & 4.31 (0.45) \\
\cmidrule(lr){2-6}

& \multirow{3}{*}{50}
  & 1 & \textbf{6.45} (0.61) & 11.5 (0.97) & 18.3 (1.66) \\
& & 2 & \textbf{1.51} (0.10) & 3.61 (0.34) & 5.74 (0.53) \\
& & 3 & \textbf{1.46} (0.02) & 2.16 (0.15) & 3.87 (0.36) \\
\cmidrule(lr){2-6}

& \multirow{3}{*}{100}
  & 1 & \textbf{6.53} (0.63) & 11.4 (0.82) & 16.4 (1.61) \\
& & 2 & \textbf{1.51} (0.08) & 3.58 (0.25) & 5.12 (0.46) \\
& & 3 & \textbf{1.46} (0.01) & 2.15 (0.12) & 3.55 (0.33) \\

\midrule
\multirow{9}{*}{2}
& \multirow{3}{*}{30}
  & 1 & \textbf{2.95} (0.25) & 3.19 (0.26) & 3.93 (0.46) \\
& & 2 & \textbf{1.65} (0.09) & 2.15 (0.15) & 3.19 (0.37) \\
& & 3 & \textbf{1.56} (0.07) & 1.83 (0.10) & 2.62 (0.27) \\
\cmidrule(lr){2-6}

& \multirow{3}{*}{50}
  & 1 & \textbf{3.05} (0.19) & 3.21 (0.23) & 3.85 (0.27) \\
& & 2 & \textbf{1.68} (0.07) & 2.15 (0.13) & 3.14 (0.25) \\
& & 3 & \textbf{1.58} (0.06) & 1.82 (0.09) & 2.55 (0.19) \\
\cmidrule(lr){2-6}

& \multirow{3}{*}{100}
  & 1 & \textbf{3.21} (0.14) & 3.25 (0.17) & 3.85 (0.23) \\
& & 2 & \textbf{1.74} (0.06) & 2.19 (0.10) & 3.10 (0.21) \\
& & 3 & \textbf{1.62} (0.05) & 1.85 (0.07) & 2.54 (0.15) \\

\bottomrule
\end{tabular}
\end{table}

\FloatBarrier

\begin{figure}[!htbp]
\centering
\includegraphics[width=\linewidth]{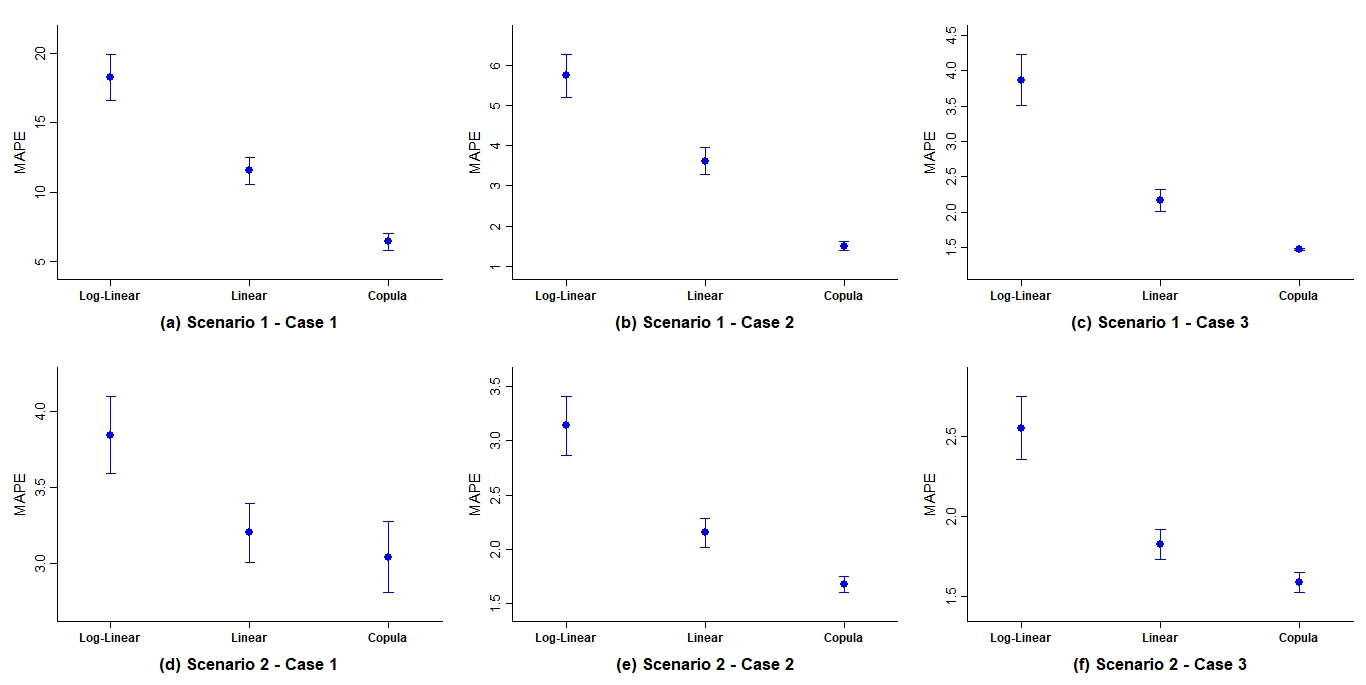}
\caption{Distribution of the mean MAPE for 500 simulations sample size n = 50 }
\label{fig:simplebox50}
\end{figure}

As illustrated in Figure~\ref{fig:simplebox50}, a significant difference is observed between the copula-based model and the log-linear model, with the copula model yielding lower MAPE values. The difference between the copula-based model and the classical linear regression model appears less pronounced; however, to formally assess this, a two-sample \textit{t}-test was conducted to compare the MAPE values of the two models across different simulation Scenarios. The hypotheses are defined as follows:
\[
\begin{aligned}
H_0 &: \text{There is no significant difference between the MAPE values of the two models.} \\
\text{vs.} \\
H_1 &: \text{There is a significant difference between the MAPE values of the two models.}
\end{aligned}
\]
For Cases 1, 2 and 3, the p-values $< 0.001$ are significantly smaller than the chosen significance level of 0.05. Therefore, the null hypothesis is rejected, indicating a statistically significant difference in MAPE between the two models. Similar patterns were observed for other sample sizes (see Figures~\ref{fig:simplebox30} and ~\ref{fig:simplebox100} in Appendix ~\ref{app:A}). 

\FloatBarrier

Overall, these results suggest that the predictive performance of the copula-based model differs significantly from that of the classical linear regression model under the considered simulation settings. More generally, the results demonstrate that copula-based regression provides a more flexible and robust predictive framework than classical linear and log-linear models in the presence of non-constant error variance and complex dependence structures.


\subsection{Study II - Multiple Predictor Copula Regression Model}
\label{sec:case1_multiple}

This study extends the simulation framework described in Section~\ref{sec:case1_single} to a multi-predictor setting with two explanatory variables. The objective is to evaluate the performance of the the proposed copula-based regression model under increased model complexity and more complex dependence structures.

The explanatory variables are generated independently, with $X_1$ following a uniform distribution on $(0,1)$ and $X_2$ following a Beta distribution with shape parameters $\alpha = 4$ and $\beta = 2$. This specification introduces heterogeneity in the marginal distributions of the covariates while maintaining bounded support.

The response variable is constructed using a linear predictor with true parameter values fixed at $\gamma_0 = 4$, $\gamma_1 = 9$, and $\gamma_2 = 6$. Random variation is incorporated through two independent error components, following the same structure described in Section~\ref{sec:case1_single}. The covariate-dependent variance component is specified as

\begin{equation}
\sigma_x^2 = k_2 \pm (a_2 X_1 + b_2 X_2)
\label{eq:10}
\end{equation}
where the constant $a_2$ and $b_2$ control the magnitude of heteroscedasticity and the shift parameter $k_2$ determines the baseline level of variance. The values of $a_2$ and $b_2$ are selected to control the degree of heteroscedasticity while preserving the underlying dependence structure between the variables. In particular, these choices are made to maintain the desired correlation levels and to reflect the different tail dependence behaviours associated with the Clayton and Gumbel copula settings. For the Gumbel case, in particular, different values are considered across the three simulation scenarios to capture varying strengths of upper-tail dependence. The simulation configurations for the different dependence levels, sample sizes, and heteroscedasticity settings are summarized in Table~\ref{tab:sim_setup_2x}.

\begin{table}[htbp]
\centering
\caption{Simulation setups for two predictor variables}
\label{tab:sim_setup_2x}
\begin{tabular}{lccccc}
\toprule
{Case} & {$\sigma$} & {$k_2$ for Clayton} & {$k_2$ for Gumble} & {$a_2$} & {$b_2$} \\
\midrule
1 & 0.20 & 0.1 & 5.5 & 4 & 1.00 \\
2 & 0.10 & 0.1 & 3.5 & 3 & 0.50 \\
3 & 0.05 & 0.1 & 2.1 & 2 & 0.01 \\
\bottomrule
\end{tabular}
\end{table}
\FloatBarrier

\begin{figure}[!htbp]
\centering
\includegraphics[width=\linewidth,height=10.5cm]{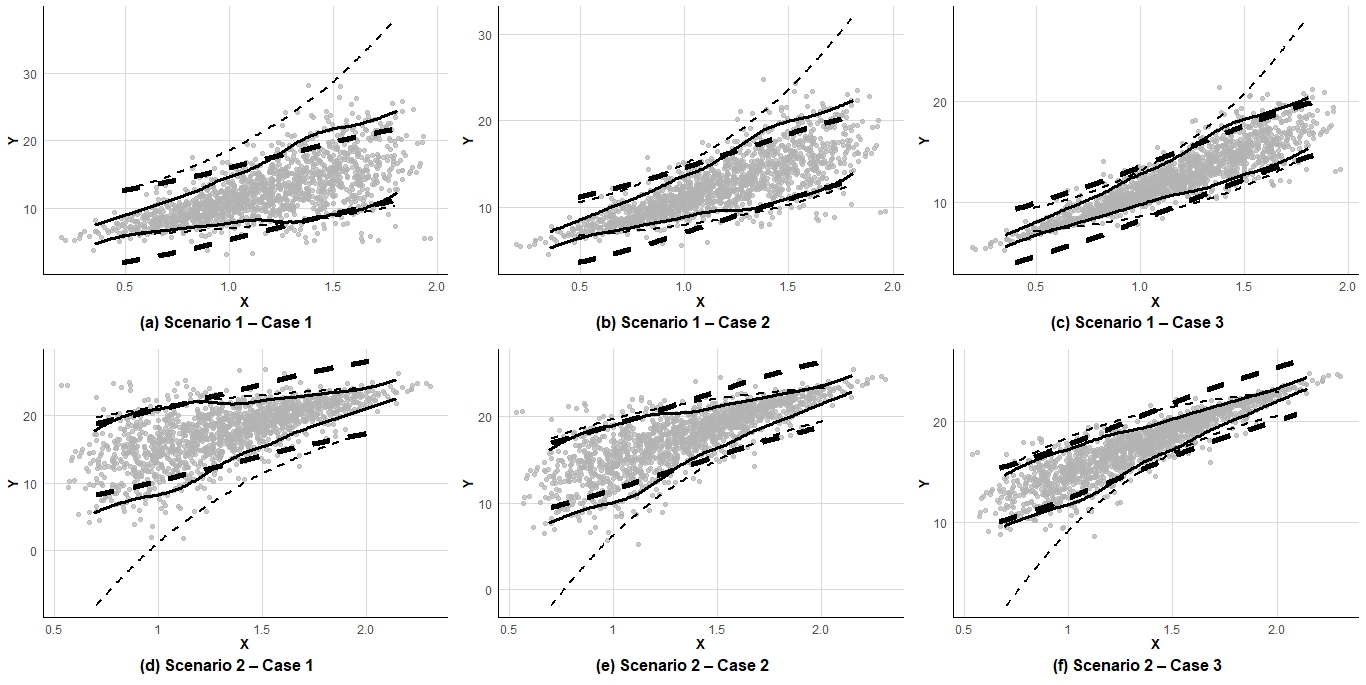}
\centering
\vspace{2mm}  
\includegraphics[width=1\linewidth,height=1.1cm,keepaspectratio]{model_nam.jpeg}

\caption{95\% prediction intervals for different correlation levels with sample size = 50 }
\label{fig:ci_all_mul_50}
\end{figure}

Figure~\ref{fig:ci_all_mul_50} illustrates the results obtained from the multiple-predictor simulation study. The findings are broadly consistent with those reported in Figure~\ref{fig:ci_all_50}, confirming the robustness of the earlier conclusions. The copula-based model continues to outperform both the linear and log-linear models in capturing the variability of the response variable and dependence structure of the response variable and the predictors.

Overall, these results provide further evidence that the copula-based framework offers a more flexible and accurate modelling strategy across varying simulation settings.
Similar patterns are observed for other sample sizes (see Figures~\ref{fig:A3_multi_30} and ~\ref{fig:A4_multi_100} in Appendix ~\ref{app:B})

Table~\ref{tab:mape_results_2} summarises the prediction errors across the simulation scenarios. The copula-based model consistently achieves the lowest MAPE among the competing approaches. Its relative improvement over the best non-copula model ranges from approximately 9\% to 36\%, demonstrating a meaningful gain in predictive accuracy. The improvement is particularly pronounced in several higher-dependence scenarios, highlighting the potential benefit of incorporating the dependence structure into the regression framework.

\FloatBarrier

\begin{table}[htbp]
\footnotesize
\centering
\caption{Comparison of Mean Absolute Percentage Error of each models for 500 simulations runs}
\label{tab:mape_results_2}
\renewcommand{\arraystretch}{1.2}
\begin{tabular}{l p{2.6cm} p{1.6cm} p{2.6cm} p{2.6cm} p{2.6cm}}
\toprule
\multirow{2}{*}{Scenario}
& \multirow{2}{*}{Sample Size} 
& \multirow{2}{*}{Case} 
& \multicolumn{3}{c}{Mean MAPE for Each Model (with standard deviation)} \\ 
\cmidrule(lr){4-6}
 &  &  
& {Copula} 
& {Linear} 
& {Log-Linear} \\ 
\midrule

\multirow{9}{*}{1}
& \multirow{3}{*}{30}
  & 1 & \textbf{0.22} (0.0302) & 0.31 (0.0307) & 0.37 (0.0602) \\
& & 2 & \textbf{0.15} (0.0206) & 0.21 (0.0217) & 0.23 (0.0258) \\
& & 3 & \textbf{0.08} (0.0106) & 0.12 (0.0110) & 0.17 (0.0115) \\
\cmidrule(lr){2-6}

& \multirow{3}{*}{50}
  & 1 & \textbf{0.22} (0.0232) & 0.30 (0.0256) & 0.35 (0.0581) \\
& & 2 & \textbf{0.14} (0.0161) & 0.20 (0.0162) & 0.23 (0.0177) \\
& & 3 & \textbf{0.08} (0.0078) & 0.12 (0.0081) & 0.17 (0.0084) \\
\cmidrule(lr){2-6}

& \multirow{3}{*}{100}
  & 1 & \textbf{0.22} (0.0175) & 0.30 (0.0221) & 0.36 (0.0722) \\
& & 2 & \textbf{0.14} (0.0122) & 0.20 (0.0142) & 0.22 (0.0151) \\
& & 3 & \textbf{0.07} (0.0065) & 0.11 (0.0067) & 0.14 (0.0068) \\

\midrule
\multirow{9}{*}{2}
& \multirow{3}{*}{30}
  & 1 & \textbf{0.67} (0.0243) & 0.74 (0.0845) & 0.96 (0.0977) \\
& & 2 & \textbf{0.15} (0.0049) & 0.21 (0.0211) & 0.34 (0.0468) \\
& & 3 & \textbf{0.09} (0.0025) & 0.14 (0.0108) & 0.24 (0.0324) \\
\cmidrule(lr){2-6}

& \multirow{3}{*}{50}
  & 1 & \textbf{0.66} (0.0172) & 0.72 (0.0638) & 0.95 (0.0926) \\
& & 2 & \textbf{0.15} (0.0045) & 0.21 (0.0137) & 0.33 (0.0462) \\
& & 3 & \textbf{0.09} (0.0022) & 0.13 (0.0071) & 0.23 (0.0278) \\
\cmidrule(lr){2-6}

& \multirow{3}{*}{100}
  & 1 & \textbf{0.66} (0.0134) & 0.70 (0.0507) & 0.87 (0.0687) \\
& & 2 & \textbf{0.15} (0.0035) & 0.21 (0.0109) & 0.31 (0.0295) \\
& & 3 & \textbf{0.09} (0.0011) & 0.13 (0.0051) & 0.21 (0.0162) \\

\bottomrule
\end{tabular}
\end{table}
\FloatBarrier

The two sample \textit{t}-test was also conducted to assess the significance of the differences in MAPE. The results are consistent with those obtained in the first simulation, with p-values less than 0.05 for all cases. Therefore, the null hypothesis is rejected in all scenarios, indicating that the differences in predictive accuracy between the copula-based model and the classical linear regression model are statistically significant. Similar patterns are observed for other sample sizes (see Figures~\ref{fig:multiplebox30} and ~\ref{fig:multiplebox100} in Appendix ~\ref{app:B}).

\begin{figure}[!htbp]
\centering
\includegraphics[width=\linewidth]{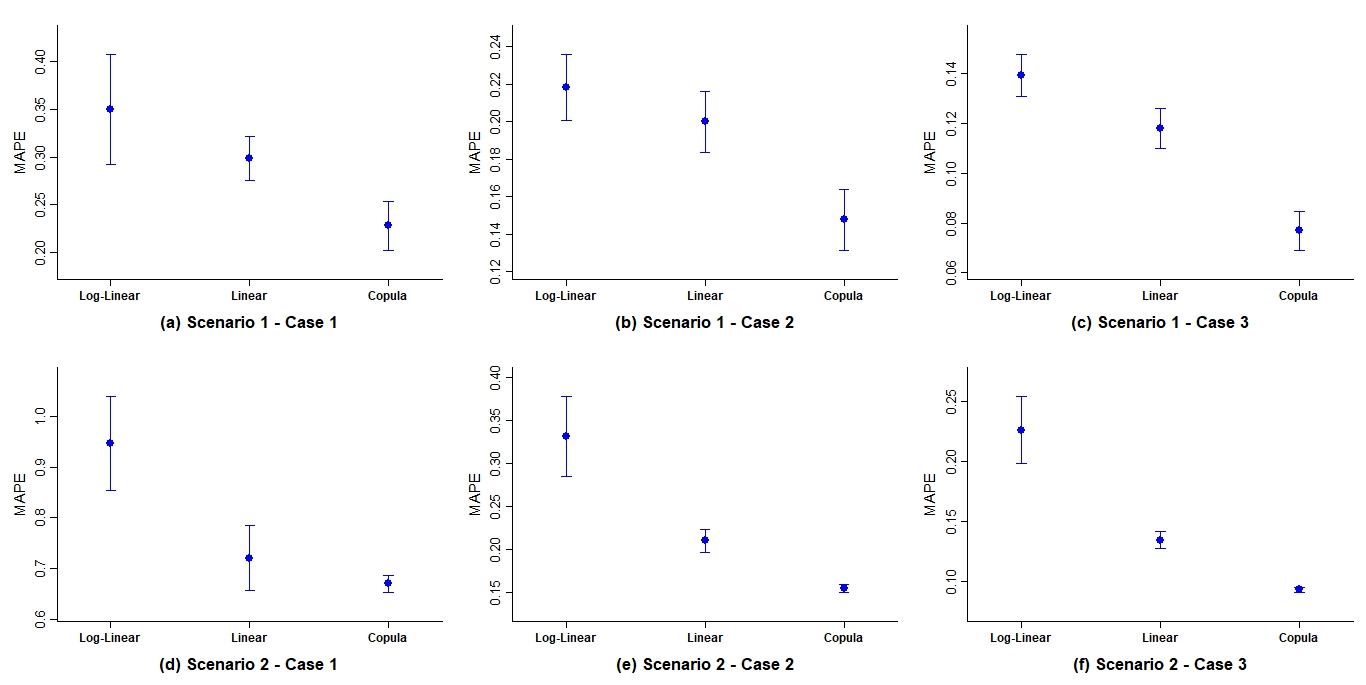}

\caption{Distribution of the mean MAPE for 500 simulations sample size n = 50 }
\label{fig:multiplebox50}
\end{figure}

\FloatBarrier
 
These findings further reinforce the robustness of the copula-based regression framework across different simulation settings, demonstrating its effectiveness in modelling complex dependence structures and non-constant error variance.

\section {Application}
\label{Application}

This section presents the applications of the proposed copula-based regression framework to two real-world datasets. The first application uses the Engel dataset available in the \texttt{quantreg} R package, which contains household food expenditure and income data and is widely used for analysing the relationship between household income and food expenditure. The second application uses the Wage dataset available in the \texttt{ISLR} R package, which contains information on individuals' earnings together with demographic and job-related characteristics, including age, education level, job class, and health status. The Breusch-Pagan test for heteroscedasticity produced $p < 0.001$, confirming the presence of non-constant error variance in both applications. These findings suggest that the assumptions of classical linear regression are violated and motivate the use of more flexible modelling approaches. 
The proposed copula-based regression model was applied to both applications and compared with classical linear and log-linear regression models as standard benchmarks, median quantile regression ($\tau=0.50$) as a robust alternative for heteroscedastic and non-normal data, and GAMLSS as a flexible distributional regression benchmark that models multiple parameters of the response distribution, including location and scale, as functions of the explanatory variables.

Initially, the datasets were preprocessed by removing observations with missing values and outliers and were randomly divided into training (70\%) and testing (30\%) sets. For the first application~\ref{sec:case2_real_dataset}, an Inverse Gaussian GAMLSS model with income-dependent scale was employed, reflecting the positive and heteroscedastic nature of food expenditure. For the second application~\ref{sec:case1_real_dataset}, Normal, Gamma, Log-Normal, and Box--Cox Power Exponential distributions were considered within the GAMLSS framework, with the final model selected based on model-selection criteria and diagnostic performance. Each model was fitted to the training data and evaluated on the corresponding testing data in both applications. This procedure was repeated 100 times using different random splits, and the resulting MAPE values were summarised by their mean, standard deviation, and 95\% prediction interval.

Following Step 2 of Algorithm~\ref{alg:copula_updated}, the Gamma distribution was selected as the marginal distribution of the linear predictor $Z$ based on the AIC and BIC values (see Figure~\ref{fig:density1} and Figure~\ref{fig:density2}). The copula regression model was then fitted with the Gamma marginal for $Z$ and an Archimedean copula to capture the dependence between $Z$ and $Y$.

\subsection{Application I: Engel Food Expenditure Data}
\label{sec:case2_real_dataset}

\begin{figure}[!htbp]
\centering
\includegraphics[width=\linewidth,height=8cm]{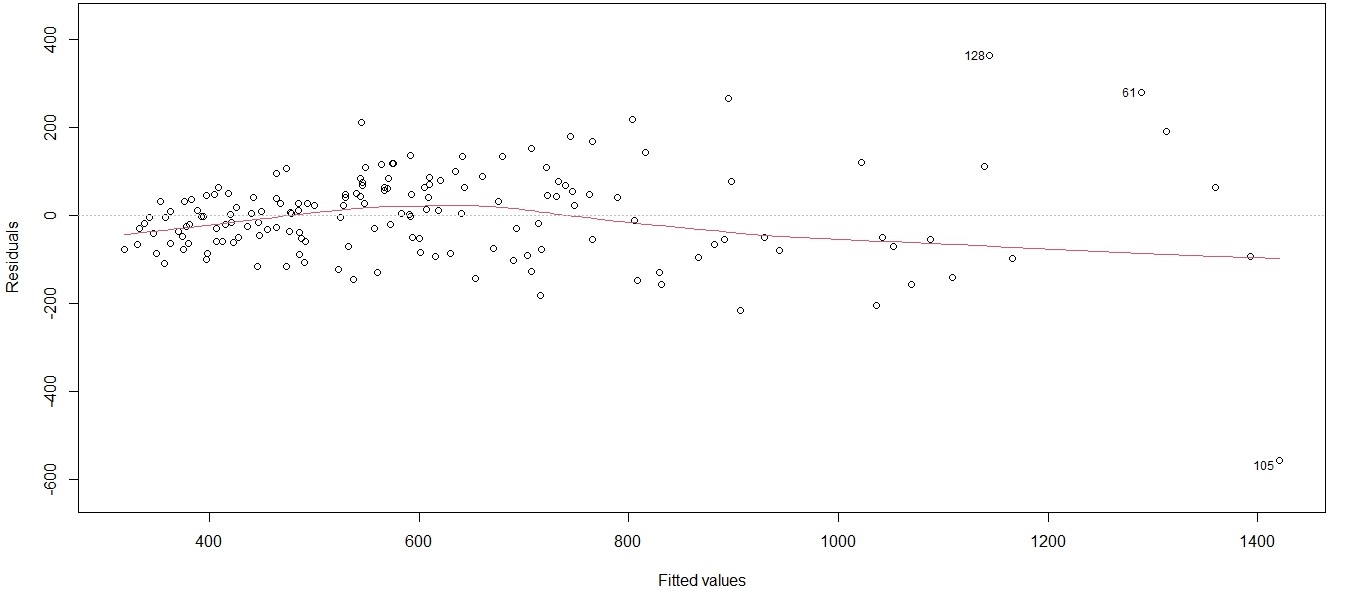}
\caption{Plot of residuals against predicted values in the linear regression model}
\label{fig:residuals_versus2}
\end{figure}

To provide a complementary evaluation of the proposed framework in a simple linear regression setting, the Engel food-expenditure data were considered. The linear model also achieved an $R^2$ value of 0.839, indicating that a substantial proportion of the variation in food expenditure is explained by household income. Figure~\ref{fig:residuals_versus2} presents the  residuals versus fitted value plot for the classical linear regression model. The residual spread increases with the fitted values, indicating heteroscedasticity. 

\FloatBarrier

\begin{figure}[!htbp]
\centering
\includegraphics[width=\linewidth,height=8cm]{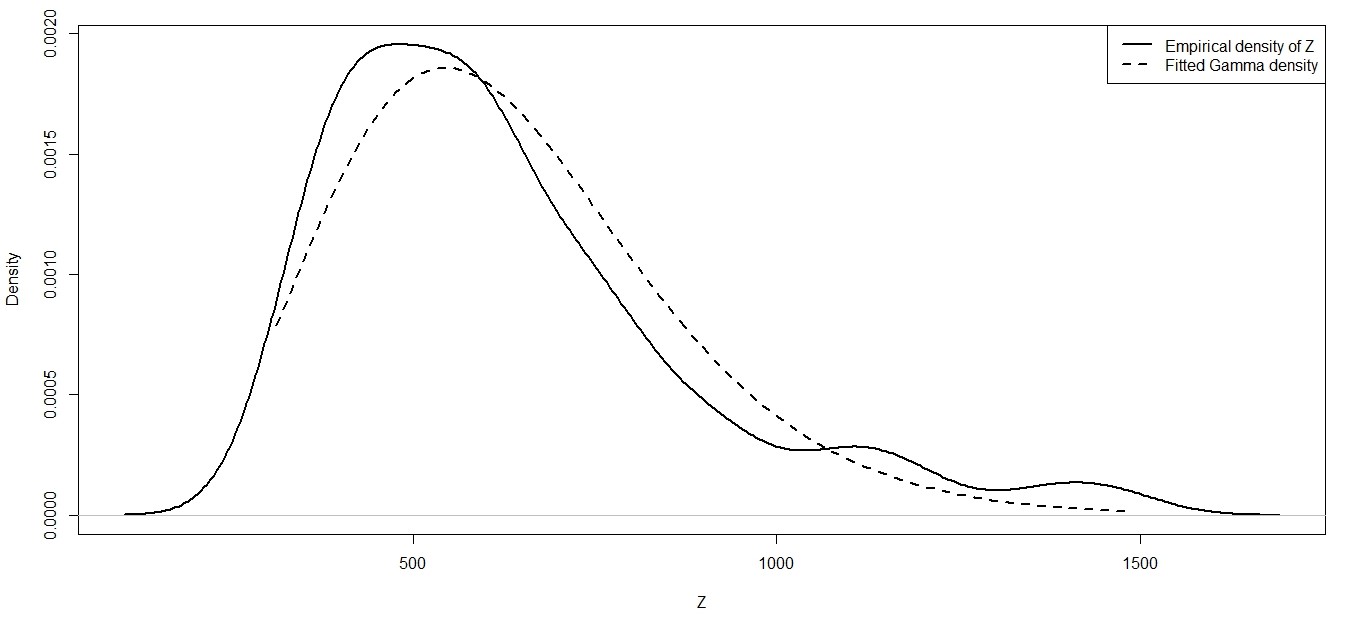}
\caption{The density curve for first application}
\label{fig:density1}
\end{figure}

Figures~\ref{fig:density1} show a close agreement between the empirical and fitted Gamma densities, confirming the adequacy of the Gamma distribution for the marginal distribution of $Z$.

\FloatBarrier

The estimated regression coefficients and standard errors are presented in Table~\ref{tab:coeffs_comparison_case2}. Although the magnitude of the estimates varies across models, the coefficients show consistent directions of association, indicating similar relationships between the predictor and response variables across the different modelling frameworks.

\begin{table}[ht]
\footnotesize
\centering
\setlength{\tabcolsep}{5pt}
\caption{Estimated regression coefficients and standard errors}
\label{tab:coeffs_comparison_case2}
\begin{tabular}{l>{\centering\arraybackslash}p{6cm}c}
\hline
Model & Intercept (with standard error) & Age (with standard error)  \\
\hline
MLR        & 122.291 (18.765) & 0.509 (0.017)   \\
Log-linear & 5.616  (0.033)  & $7.526\times10^{-4}$ ($3.065\times10^{-5}$)  \\
GAMLSS     & 5.491  (0.039)  & $9.094\times10^{-4}$ ($4.459\times10^{-5}$)  \\
QR         & 102.900 (32.552) & 0.541 (0.045)  \\
Copula     & 480.541 (22.823) & 0.127 (0.024)  \\
\hline
\end{tabular}
\end{table}
\FloatBarrier

The predictive performance of the five competing models over 100 simulation runs is summarised in Table~\ref{tab:model1_results}. The proposed copula-based model achieves the lowest mean MAPE for both training and testing data, followed by GAMLSS and MLR. It also shows relatively low variability and a narrow 95\% prediction interval, indicating stable predictive performance. 

\begin{table}[ht]
\centering
\caption{Training and testing prediction performance of each models for 100 simulations runs}
\begin{tabular}{lccc ccc}
\hline
\multirow{2}{*}{Model} 
& \multicolumn{3}{c}{Training} & \multicolumn{3}{c}{Testing} \\
\cline{2-4} \cline{5-7}
& Mean MAPE & SD & 95\% Intervals & Mean MAPE & SD & 95\% Intervals \\
\hline     
MLR        & 0.116 & 0.469 & (0.106, 0.123)
           & 0.117 & 0.937 & (0.101, 0.134) \\
Log-linear & 0.144 & 0.534 & (0.133, 0.154)
           & 0.147 & 1.373 & (0.124, 0.174) \\
GAMLSS     & 0.113 & 0.412 & (0.106, 0.121)
           & 0.116 & 1.005 & (0.099, 0.138) \\
QR         & 0.155 & 0.739 & (0.140, 0.168)
           & 0.159 & 2.075 & (0.127, 0.196) \\
Copula     & 0.112 & 0.395 & (0.105, 0.119)
           & 0.113 & 0.948 & (0.097, 0.132) \\
\hline
\end{tabular}
\label{tab:model1_results}
\end{table}
\FloatBarrier

\subsection{Application II: Wage Data}
\label{sec:case1_real_dataset}

\begin{figure}[!htbp]
\centering
\includegraphics[width=\linewidth,height=8cm]{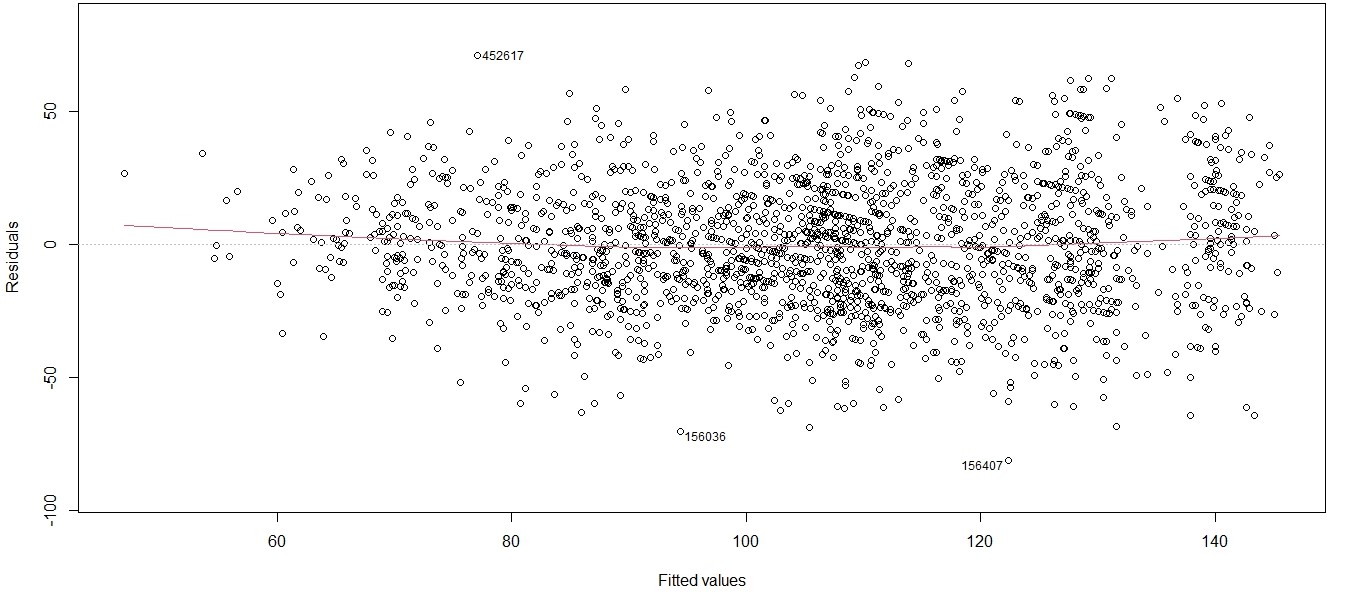}
\caption{Plot of residuals against predicted values in the linear regression model}
\label{fig:residuals_versus}
\end{figure}

Figure~\ref{fig:residuals_versus} presents the residuals versus fitted values plot for the linear regression model. Although the residuals are centred around zero, their variability increases with larger fitted values, indicating a clear heteroscedastic pattern. Although the classical linear regression model explains approximately 41\% of the variation in wages $R^2=0.408$, a substantial proportion of the variability remains unexplained. 

\FloatBarrier

\begin{figure}[!htbp]
\centering
\includegraphics[width=\linewidth,height=8cm]{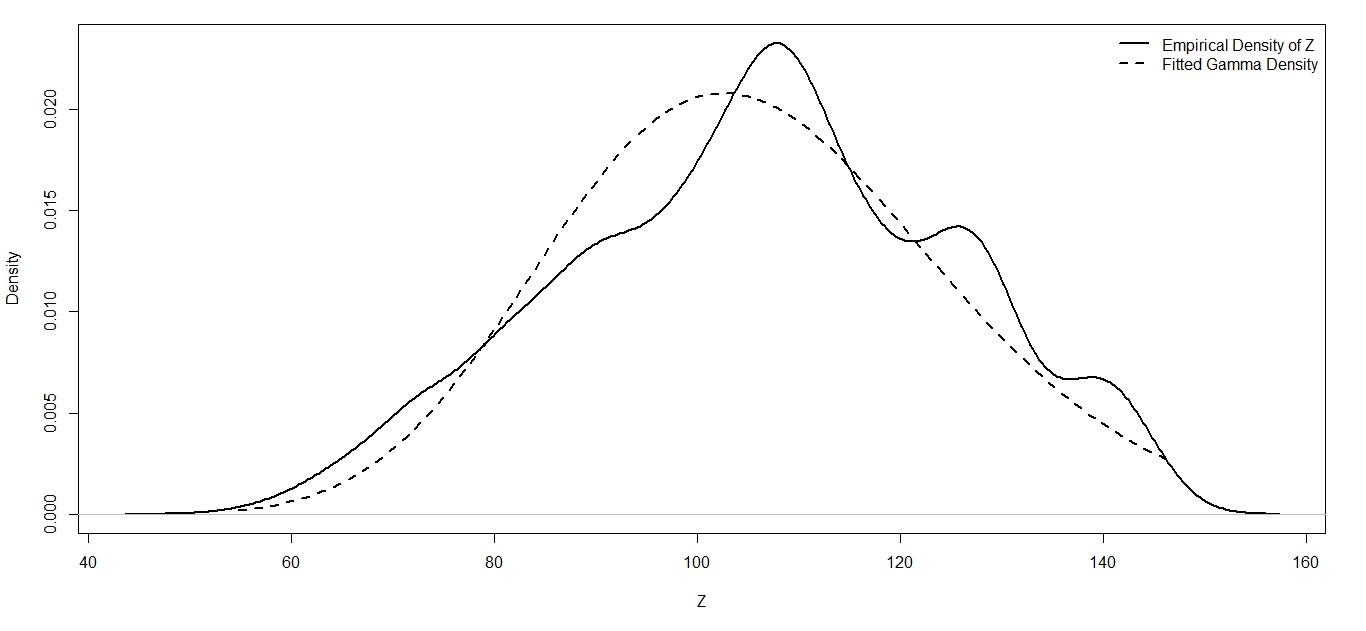}
\caption{The density curve for second application}
\label{fig:density2}
\end{figure}

Figures~\ref{fig:density2} show a close agreement between the empirical and fitted Gamma densities, confirming the adequacy of the Gamma distribution for the marginal distribution of $Z$.

\FloatBarrier

Table~\ref{tab:coeffs_comparison} presents the estimated regression coefficients and standard errors across the five models. Despite differences in coefficient magnitude, the models generally show consistent directions of association. In particular, age and higher education levels have positive associations with the response, while Black and Other racial groups and those without health insurance show negative associations. Overall, the competing models identify similar broad patterns of relationships across the explanatory variables.

\begin{table}[ht]
\footnotesize
\centering
\setlength{\tabcolsep}{3pt} 
\caption{Estimated regression coefficients and standard errors}
\label{tab:coeffs_comparison}
\begin{tabular}{lrrrrrrrrrr}
\hline
Coefficients & \multicolumn{2}{c}{MLR} & \multicolumn{2}{c}{Log‑linear} & \multicolumn{2}{c}{GAMLSS} & \multicolumn{2}{c}{QR} & \multicolumn{2}{c}{Copula}\\
 & Estimate & SE & Estimate & SE & Estimate & SE & Estimate & SE & Estimate & SE \\
\hline
intercept                     & 68.186  &  3.017 & 4.232 & 0.033 & 68.743  &  2.553  & 64.217 & 3.171 & 82.183 & 1.658 \\
age                           &  0.217  &  0.053 & 0.002 & $5.00 \times 10^{-4}$ &  0.215  &  0.051  & 0.254 & 0.060 & 0.094 & 0.038 \\
maritl2. Married              & 15.545  &  1.457 & 0.157 & 0.015 & 14.666  &  1.274  & 15.046 & 1.492 & 6.576 & 0.852 \\
maritl3. Widowed              & 5.125  &  6.244 & 0.058 & 0.064 &  0.837  &  5.251  & 12.416 & 2.157 & 4.535 & 4.542 \\
maritl4. Divorced             & 5.839  &  2.444 & 0.057 & 0.025 & 5.838  &  2.189  & 6.673 & 2.853 & 2.197 & 1.759 \\
maritl5. Separated            & 10.723  &  3.905 & 0.116 & 0.039  & 8.169  &  3.312  & 12.448 & 6.830 & 6.499 & 2.789 \\
race2. Black                  & -3.763  &  1.822 & -0.026 & 0.019 & -2.459  &  1.585 & -4.436 & 1.858 & -1.663 & 1.309 \\
race3. Asian                  & 1.114  &  2.191 & -0.012 & 0.022 & -0.787  &  2.151  & 0.861 & 2.918 & 0.553 & 1.636 \\
race4. Other                  & -9.131  &  5.067 & -0.085 & 0.052 & -5.900  &  3.769  & -9.086 & 7.518 & -3.357 & 3.525 \\
edu2. HS Grad           &  8.682  &  2.026 & 0.099 & 0.021 &  8.502  &  1.534  & 10.960 & 1.719 & 4.684 & 1.346 \\
edu3. Some College      & 18.141  &  2.157 & 0.196 & 0.022 & 17.344  &  1.729  & 21.061 & 1.961 & 8.741 & 1.297 \\
edu4. College Grad      & 28.507  &  2.188 & 0.281 & 0.022 & 29.067  &  1.895  & 32.178 & 2.494 & 13.257 & 0.958 \\
edu5. Adv Degree   & 40.689  &  2.444 & 0.379 & 0.025 & 40.163  &  2.365  & 44.395 & 2.624 & 19.234 & 0.521 \\
jobclass2. Information        & 0.000  &  0.000 & 0.018 & 0.011 &  1.387  &  1.042  & 0.912 & 1.303 & 0.803 & 0.815 \\
health2.>=VeryGood          &  5.653  &  1.186 & 0.059 & 0.012 &  4.722  &  1.052 & 4.116 & 1.372 & 2.756 & 0.824 \\
health\_ins2. No              & -17.485 & 1.172 & -0.195 & 0.012 & -14.875  &  1.059  & -15.915 & 1.342 & -6.700 & 0.549 \\
\hline
\end{tabular}
\end{table}
\FloatBarrier
                   
The training results assessed model fit, while the testing results evaluated predictive performance. As shown in Table~\ref{tab:model2_results}, all five models achieved similar performance for the Wage data. The log-linear model achieved the lowest testing MAPE, while the copula model produced comparable predictive performance. Overall, no substantial performance differences were observed among the models.

\begin{table}[ht]
\centering
\caption{Training and testing prediction performance of each models for 100 simulations runs}
\begin{tabular}{lccc ccc}
\hline
\multirow{2}{*}{Model} 
& \multicolumn{3}{c}{Training} & \multicolumn{3}{c}{Testing} \\
\cline{2-4} \cline{5-7}
& Mean MAPE & SD & 95\% Intervals & Mean MAPE & SD & 95\% Intervals \\
\hline     
MLR        & 0.205 & 0.271 & (0.199, 0.210)
           & 0.205 & 0.683 & (0.192, 0.219) \\
Log-linear & 0.199 & 0.258 & (0.194, 0.204)
           & 0.199 & 0.647 & (0.186, 0.212) \\
GAMLSS     & 0.203 & 0.281 & (0.197, 0.209)
           & 0.204 & 0.698 & (0.191, 0.218) \\
QR         & 0.204 & 0.284 & (0.199, 0.210)
           & 0.205 & 0.695 & (0.190, 0.219) \\
Copula     & 0.208 & 0.289 & (0.201, 0.214)
           & 0.207 & 0.626 & (0.202, 0.210) \\
\hline
\end{tabular}
\label{tab:model2_results}
\end{table}
\FloatBarrier

Overall, the two applications demonstrate that predictive performance depends on the characteristics of the data and the criterion considered. In the first application, the relatively high $R^2$ indicates a strong linear association; however, the presence of heteroscedasticity means that $R^2$ alone is insufficient to establish model adequacy. In the second application, the linear model provides a less satisfactory representation, while the copula-based model achieves the lowest MAPE among the competing approaches. Nevertheless, the copula model remains competitive in the first application as well. These findings suggest that the proposed copula-based regression provides a flexible and competitive alternative for prediction when heteroscedasticity is present, particularly when predictive accuracy and dependence structure are of interest.

\section{Conclusion}
\label{Conclusion}

This study addresses the limitations of classical regression approaches for modelling relationships in the presence of heteroscedasticity. Classical linear regression assumes constant error variance, while log-linear regression attempts to stabilise the variance through a logarithmic transformation. However, these approaches often fail when the error variance changes systematically with the predictors or when the dependence structure between variables is non-linear, asymmetric, or exhibits tail dependence. These limitations can reduce predictive accuracy and motivate the need for more flexible modelling approaches.

To overcome these challenges, this study proposed a copula-based regression framework that models the joint distribution of the response and explanatory variables through copula functions, enabling flexible modelling of non-linear, asymmetric, and tail-dependent relationships. The simulation results from both Simulation Studies I and II demonstrate that the proposed copula-based model consistently provides the best predictive performance across the considered two Scenarios, with its advantages being particularly evident under strong dependence and pronounced heteroscedasticity. The classical linear regression model generally performed better than the log-linear model, suggesting that the log-transformation did not consistently improve predictive performance under the simulated heteroscedastic settings.

Two real-data applications were considered to assess the practical performance of the proposed framework. The first application exhibited a relatively strong linear relationship and good predictive performance across the competing models. In contrast, the second application showed a weaker relationship and lower overall predictive performance, where the copula-based model also achieved results similar to those of the competing models. These findings suggest that the proposed model provides a competitive alternative across different data settings.

From a methodological perspective, this study develops a distribution-based modelling approach that can be applied under heteroscedastic conditions and provides consistent predictive performance. From a practical perspective, the findings highlight the importance of selecting modelling approaches according to the characteristics of the data, particularly when non-constant error variance.

Building on this work, future research can extend the proposed copula-based regression framework to discrete response settings, particularly count data applications. Copula-based approaches provide a flexible mechanism for modelling dependence while accommodating different marginal distributions. Beyond the classical Poisson and negative binomial models, recent developments have introduced more flexible distributions for handling both underdispersed and overdispersed count data, highlighting the need for more adaptable count-data methodologies \citep{Doukhan2017,SafariKatesari2020,Baker2026}. Integrating these count-data models within the proposed copula-based framework, while accounting for heteroscedasticity and dispersion, represents a promising direction for future research and would further broaden the applicability of the framework to real-world count data \citep{Pang2022}. Overall, the findings suggest that copula-based regression provides a flexible alternative to classical and log-transformed regression models for heteroscedastic data, particularly when non-normality and complex dependence structures are present.


\bibliographystyle{apalike}
\bibliography{references}

\clearpage
\appendix

\setcounter{figure}{0}
\renewcommand{\thefigure}{A\arabic{figure}}

\section{Appendix}
\label{app:A}
\subsection{Prediction Results for Simulation Study I}

\begin{figure}[!htbp]
\centering
\includegraphics[width=\linewidth,height=10.5cm]{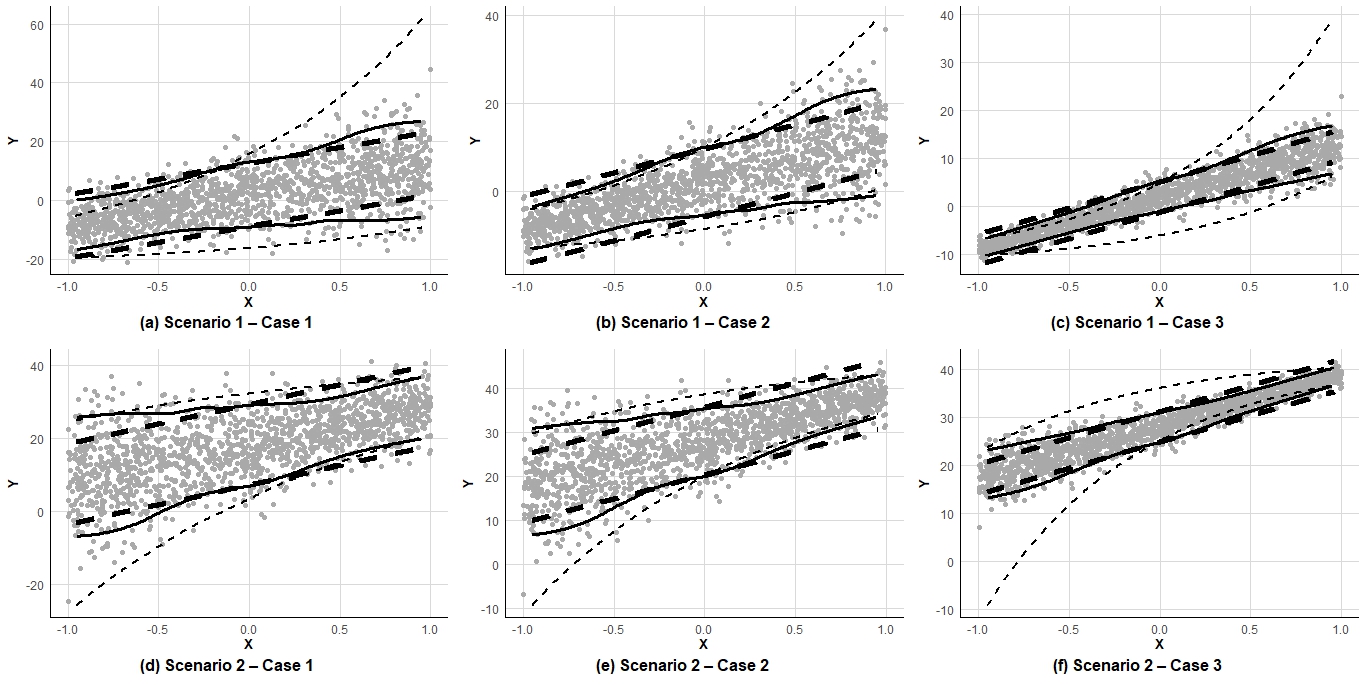}

\vspace{2mm}
\includegraphics[width=1\linewidth,height=1.1cm,keepaspectratio]{model_nam.jpeg}

\caption{95\% prediction intervals for different correlation levels with sample size $n=30$.}
\label{fig:A1_simple_30}
\end{figure}

\FloatBarrier

\begin{figure}[!htbp]
\centering
\includegraphics[width=\linewidth,height=10.5cm]{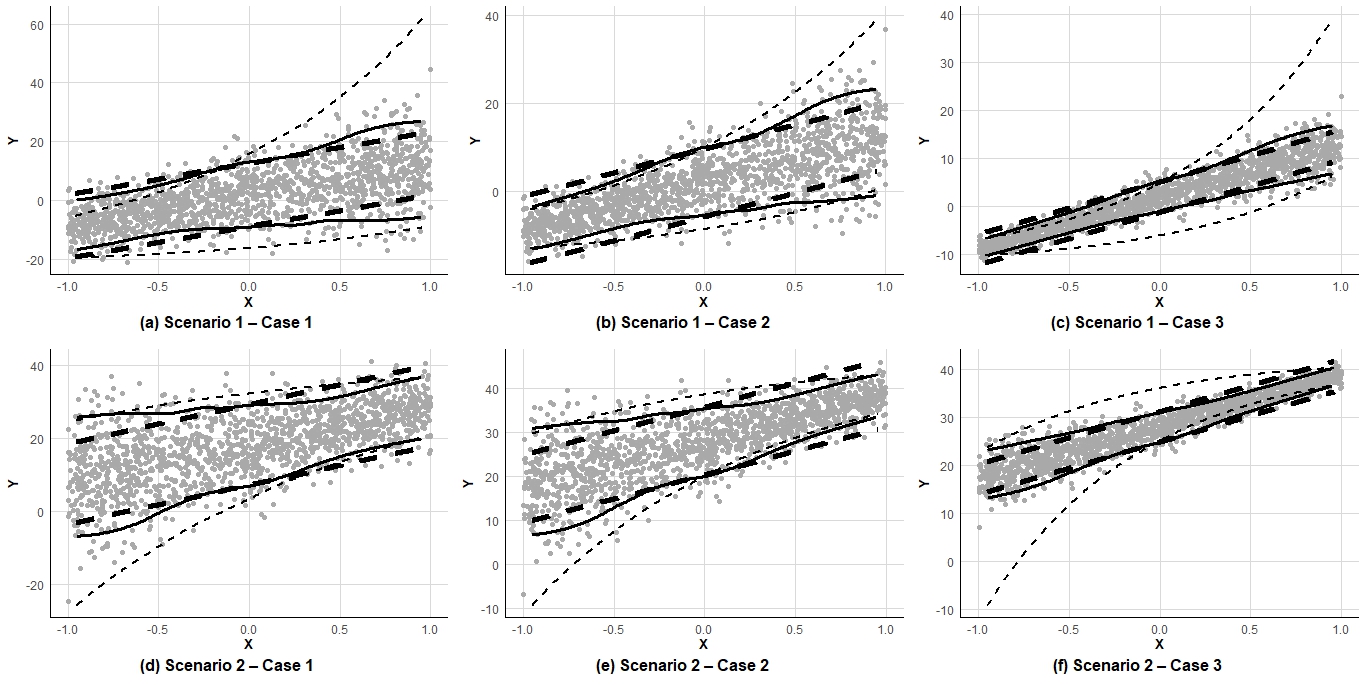}

\vspace{2mm}
\includegraphics[width=1\linewidth,height=1.1cm,keepaspectratio]{model_nam.jpeg}

\caption{95\% prediction intervals for different correlation levels with sample size $n=100$.}
\label{fig:A2_simple_100}
\end{figure}

\FloatBarrier

\begin{figure}[!htbp]
\centering
\includegraphics[width=\linewidth]{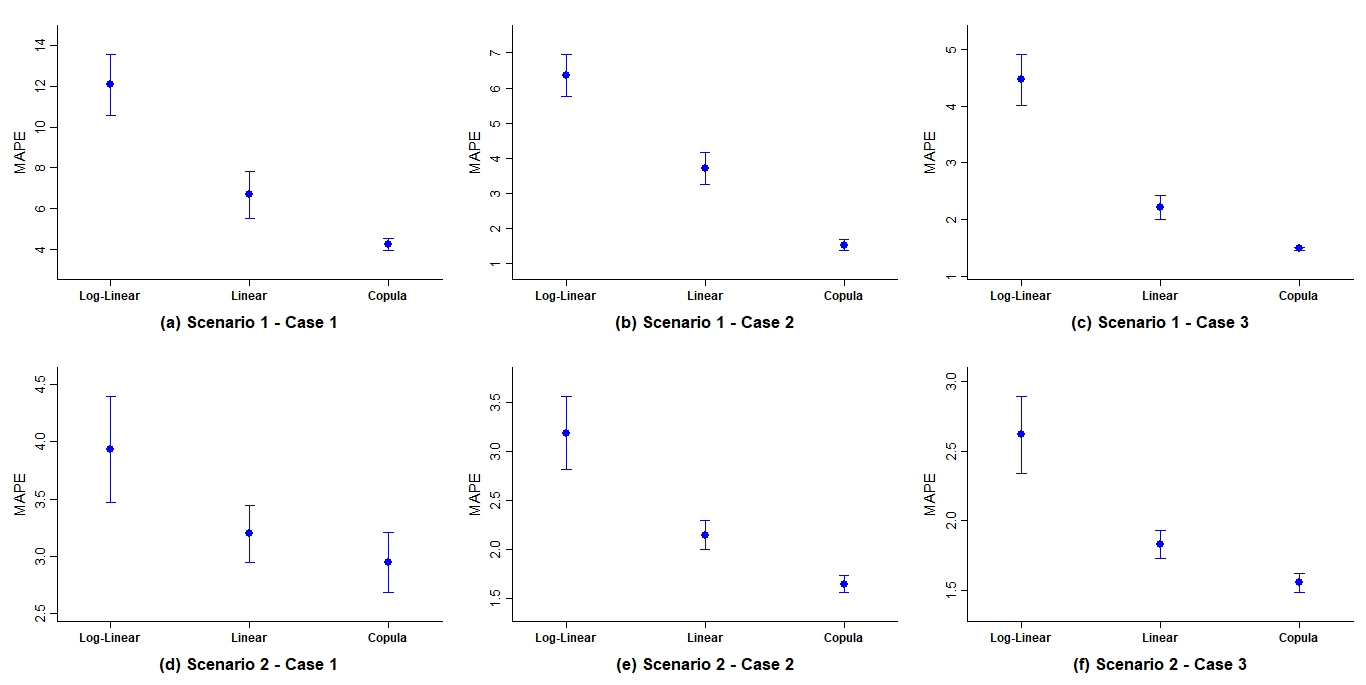}

\caption{Distribution of the mean MAPE for 500 simulations sample size n = 30 }
\label{fig:simplebox30}
\end{figure}

\FloatBarrier

\begin{figure}[!htbp]
\centering
\includegraphics[width=\linewidth]{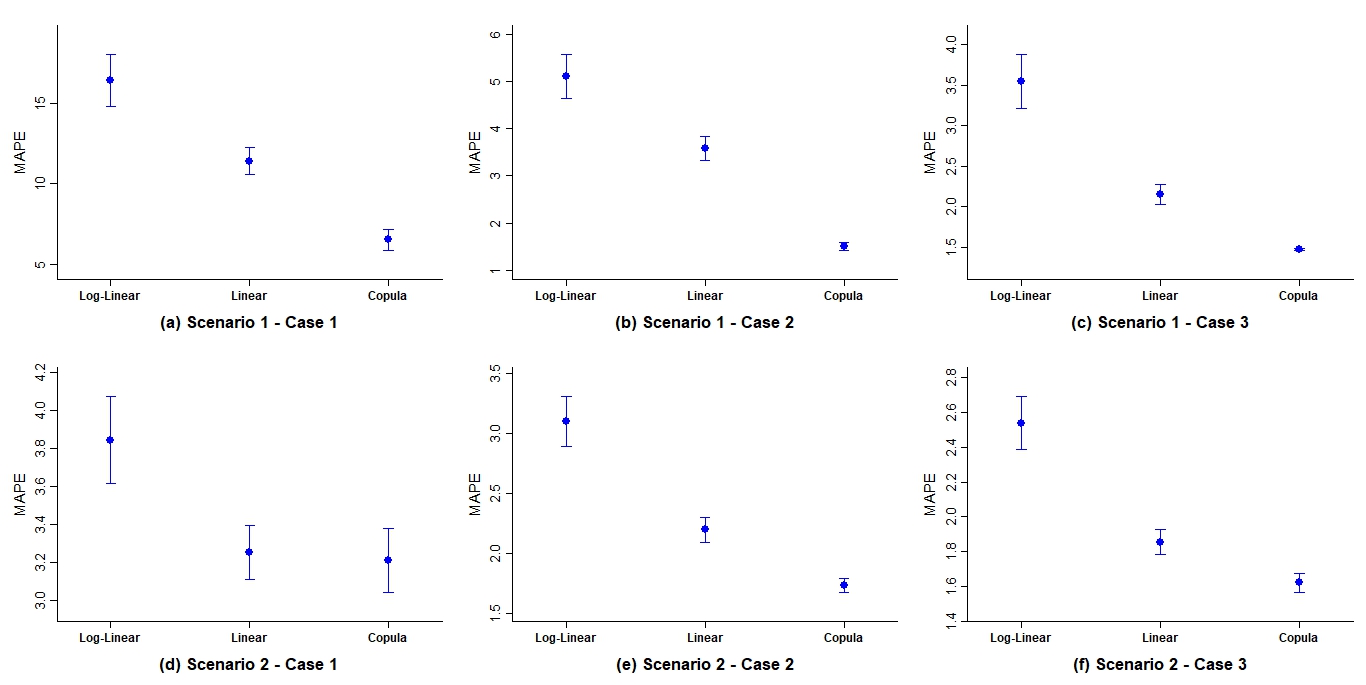}

\caption{Distribution of the mean MAPE for 500 simulations sample size n = 100 }
\label{fig:simplebox100}
\end{figure}

\FloatBarrier

\setcounter{figure}{0}
\renewcommand{\thefigure}{B\arabic{figure}}

\section{Appendix}
\label{app:B}
\subsection{Prediction Results for Simulation Study II}

\begin{figure}[!htbp]
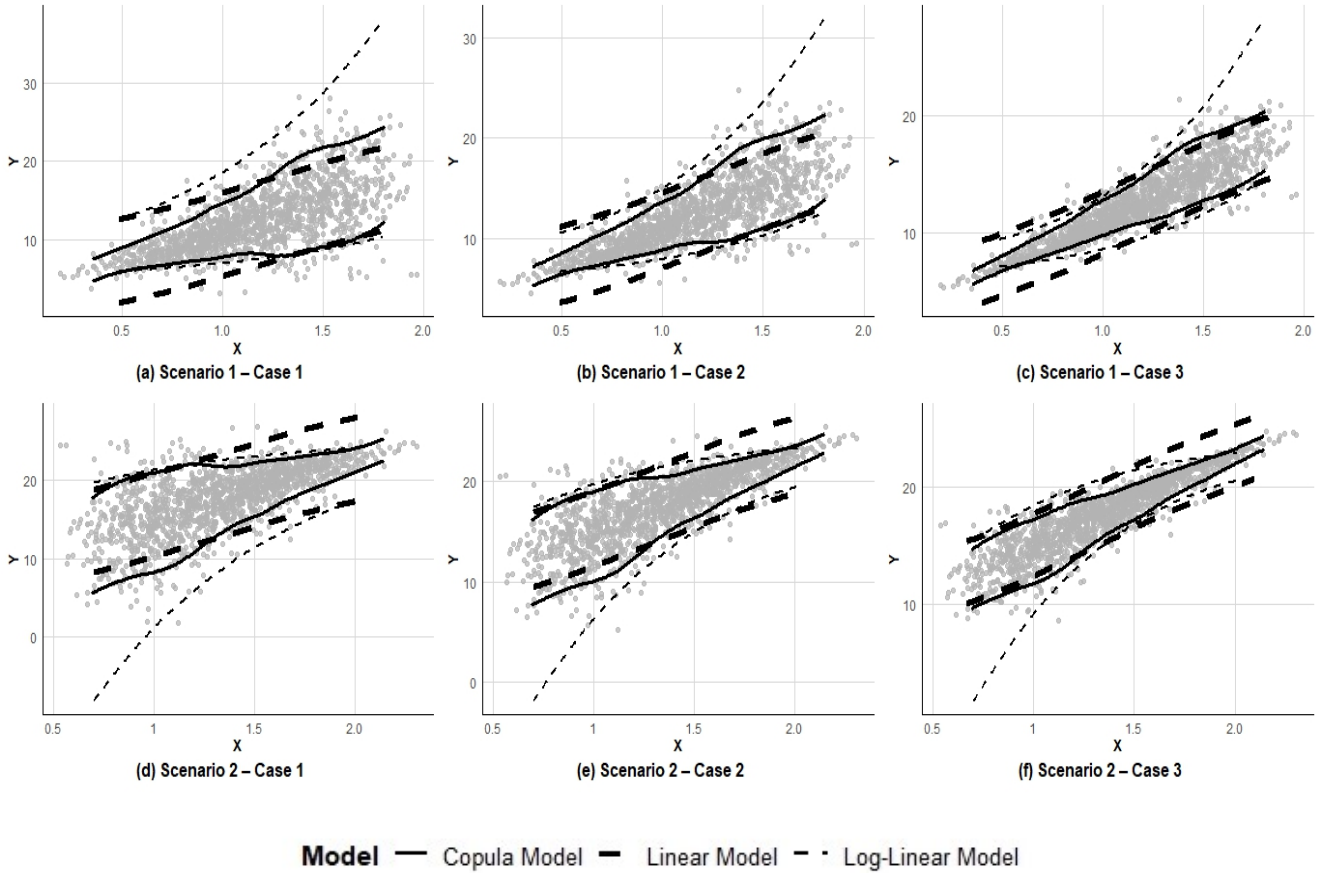

\centering
\includegraphics[width=\linewidth,height=10.5cm]{mutiple-30.jpeg}

\vspace{2mm}
\includegraphics[width=1\linewidth,height=1.1cm,keepaspectratio]{model_nam.jpeg}

\caption{95\% prediction intervals for different correlation levels with sample size $n=30$.}
\label{fig:A3_multi_30}
\end{figure}

\FloatBarrier

\begin{figure}[!htbp]
\centering
\includegraphics[width=\linewidth,height=10.5cm]{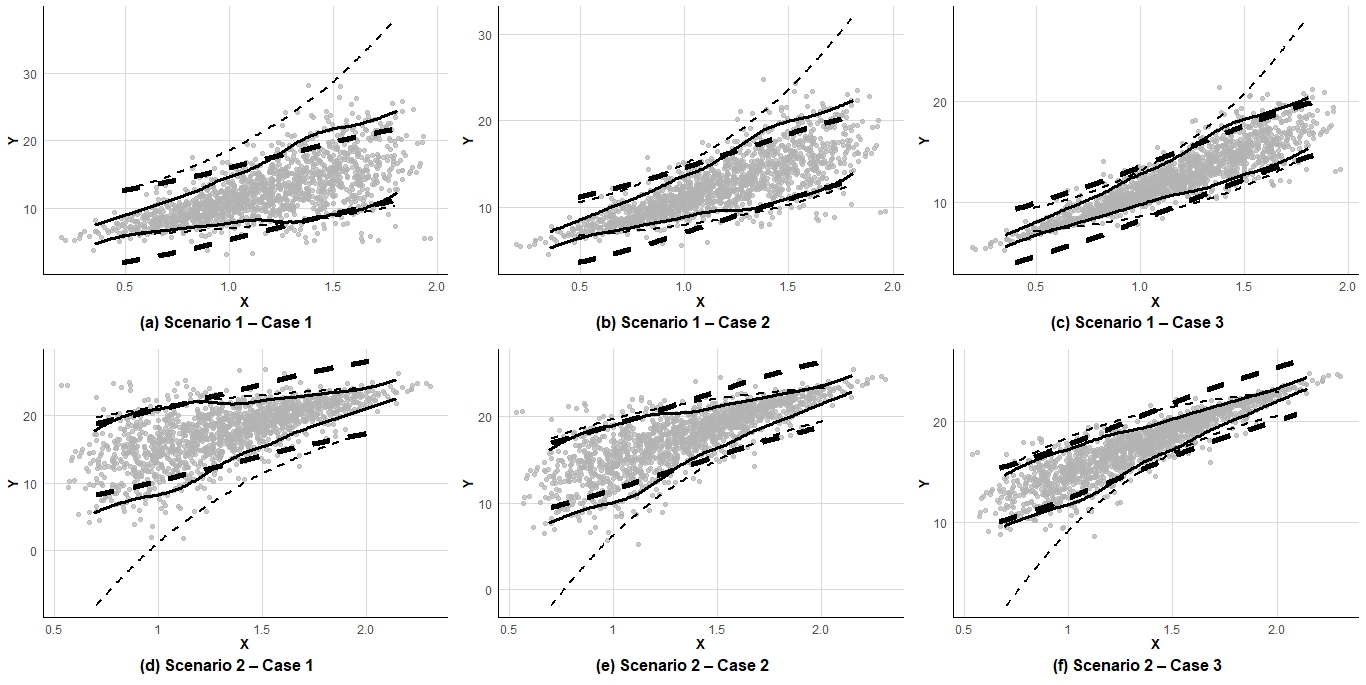}

\vspace{2mm}
\includegraphics[width=1\linewidth,height=1.1cm,keepaspectratio]{model_nam.jpeg}

\caption{95\% prediction intervals for different correlation levels with sample size $n=100$.}
\label{fig:A4_multi_100}
\end{figure}

\FloatBarrier

\begin{figure}[!htbp]
\centering
\includegraphics[width=\linewidth]{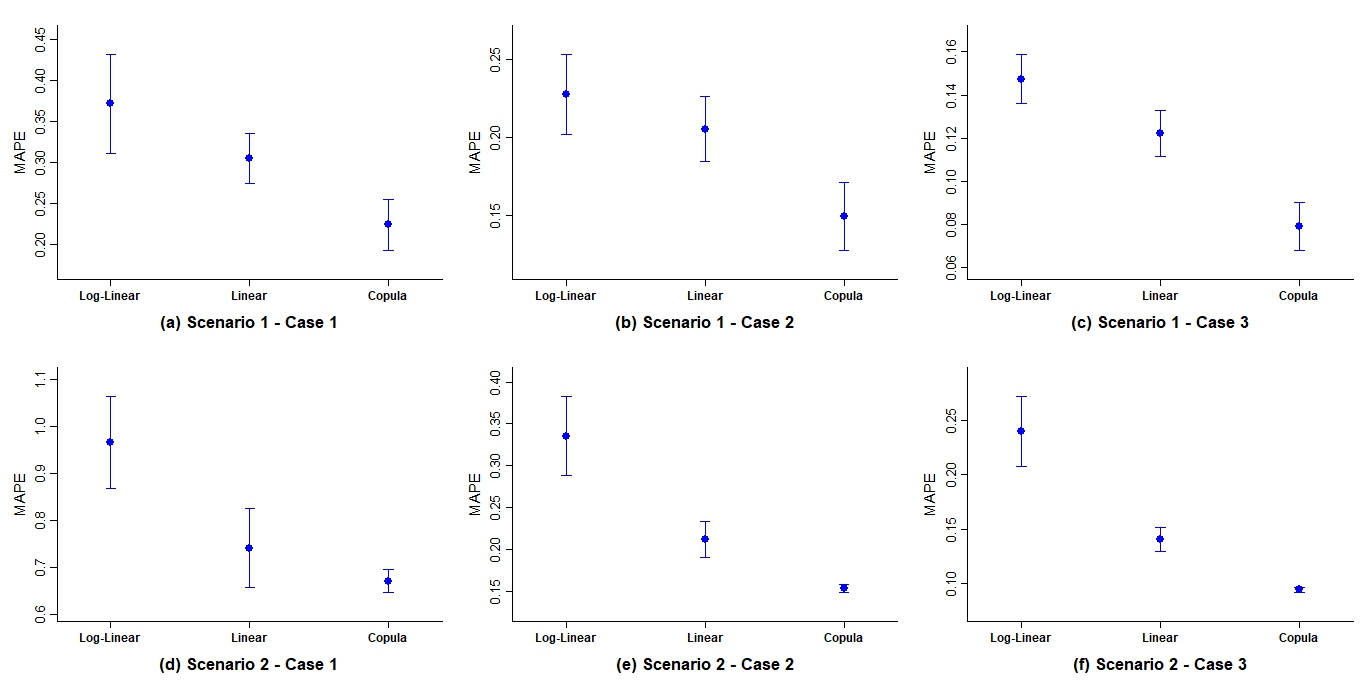}

\caption{Distribution of the mean MAPE for 500 simulations sample size n = 30 }
\label{fig:multiplebox30}
\end{figure}

\FloatBarrier

\begin{figure}[!htbp]
\centering
\includegraphics[width=\linewidth]{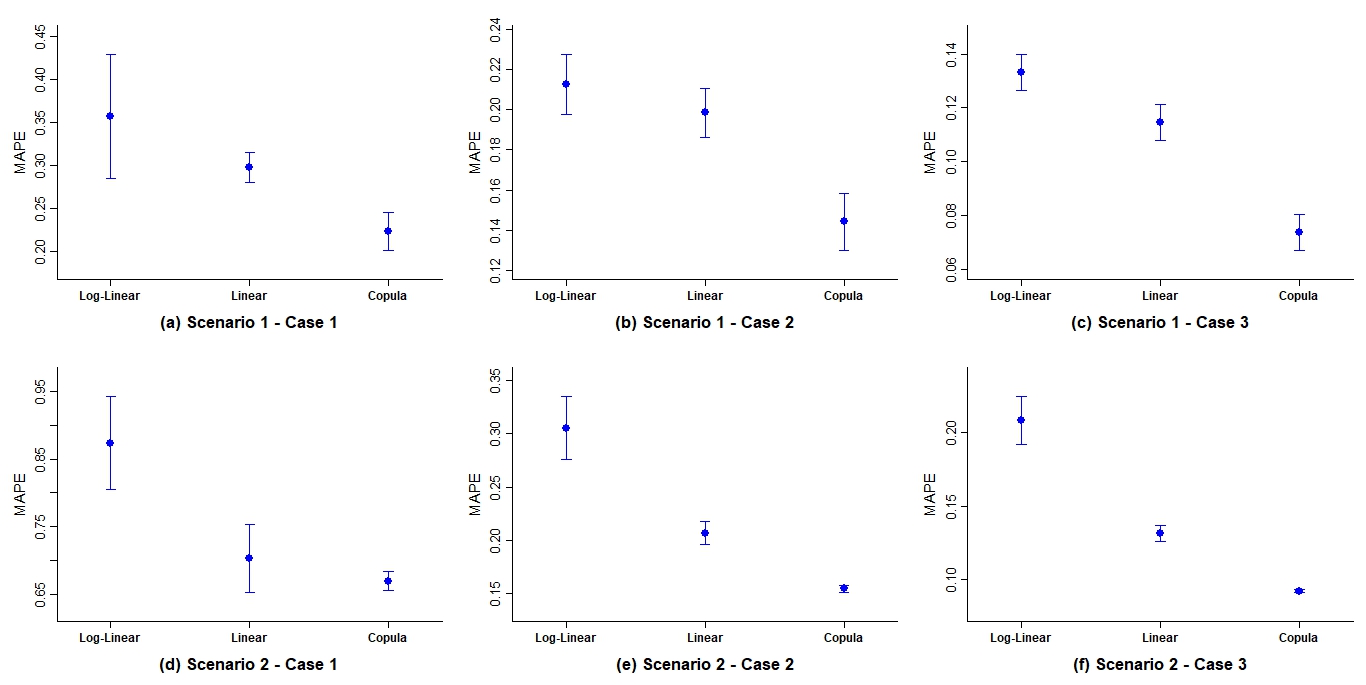}

\caption{Distribution of the mean MAPE for 500 simulations sample size n = 100 }
\label{fig:multiplebox100}
\end{figure}

\FloatBarrier

\setcounter{figure}{0}
\renewcommand{\thetable}{C\arabic{table}}
\renewcommand{\thefigure}{C\arabic{figure}}

\section{Appendix}
\label{app:C} 
\subsection{Prediction Results for Negative X}

\begin{table}[!htbp]
\centering
\caption{Comparison of Mean Absolute Percentage Error of each models for 500 simulations runs}
\label{tab:mape_results_negative}
\renewcommand{\arraystretch}{1.2}
\begin{tabular}{l p{2.6cm} p{1.6cm} p{2.6cm} p{2.6cm} p{2.6cm}}
\toprule
\multirow{2}{*}{Scenario} 
& \multirow{2}{*}{Sample Size} 
& \multirow{2}{*}{Case} 
& \multicolumn{3}{c}{Mean MAPE for Each Model(with standard deviation)} \\ 
\cmidrule(lr){4-6}
 &  &  
& Copula
& Linear
& Log-Linear \\ 
\midrule

\multirow{9}{*}{Gumble}
& \multirow{3}{*}{30}
  & 1 & \textbf{4.74} (0.53) & 7.35 (0.55) & 10.3 (0.88) \\
& & 2 & \textbf{4.69} (0.55) & 7.34 (0.58) & 9.51 (0.98) \\
& & 3 & \textbf{1.85} (0.22) & 2.51 (0.48) & 4.80 (0.60) \\
\cmidrule(lr){2-6}

& \multirow{3}{*}{50}
  & 1 & \textbf{4.59} (0.40) & 7.25 (0.41) & 9.33 (0.92) \\
& & 2 & \textbf{4.67} (0.39) & 7.35 (0.44) & 8.57 (0.83) \\
& & 3 & \textbf{1.82} (0.17) & 2.50 (0.37) & 4.12 (0.49) \\
\cmidrule(lr){2-6}

& \multirow{3}{*}{100}
  & 1 & \textbf{4.51} (0.30) & 7.21 (0.32) & 8.25 (0.94) \\
& & 2 & \textbf{4.59} (0.28) & 7.29 (0.33) & 7.67 (1.24) \\
& & 3 & \textbf{1.80} (0.13) & 2.49 (0.29) & 3.56 (0.43) \\

\bottomrule
\end{tabular}
\end{table}

\FloatBarrier

\begin{figure}[!htbp]
\centering
\includegraphics[width=0.8\linewidth,height=5.5cm]{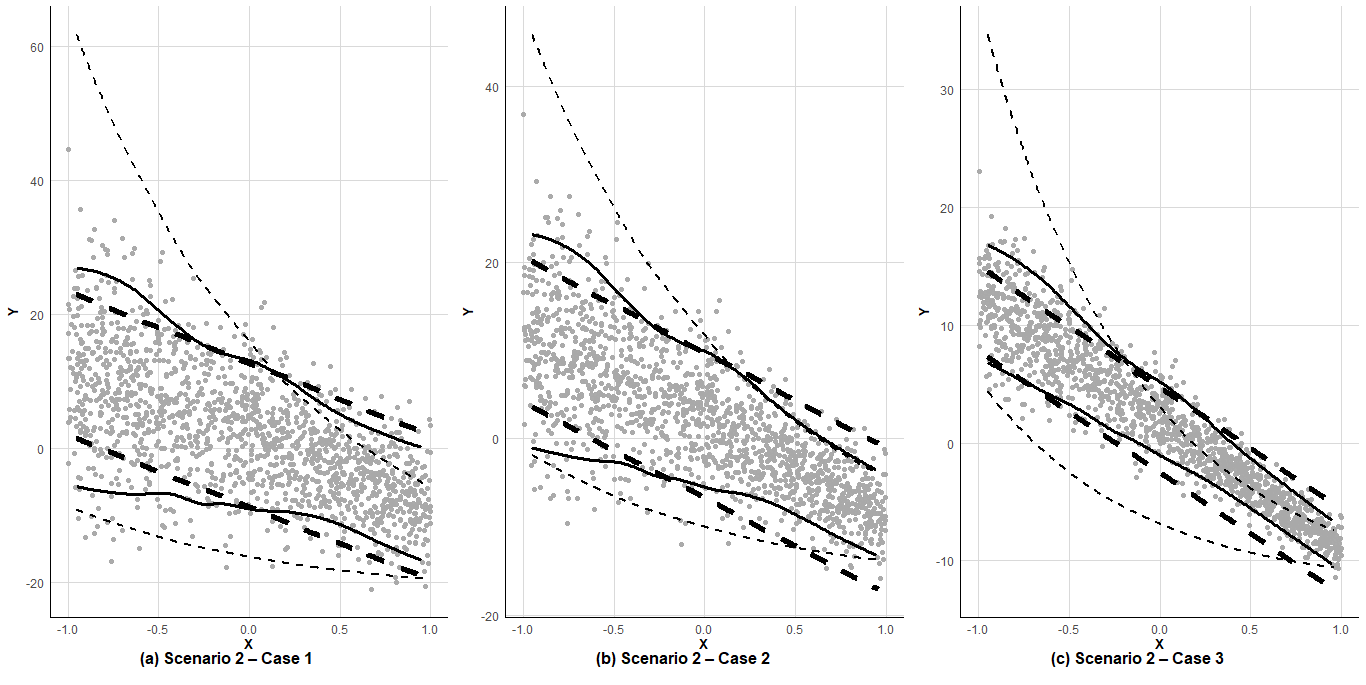}

\vspace{2mm}
\includegraphics[width=1\linewidth,height=1.1cm,keepaspectratio]{model_nam.jpeg}

\caption{95\% prediction intervals for different correlation levels with sample size $n=50$.}
\label{fig:C4_multi_100}
\end{figure}

\FloatBarrier

\end{document}